\documentclass[a4paper,11pt]{article}
\usepackage{jcappub} 
\usepackage{lineno}
\usepackage{siunitx}
\usepackage{comment}
\usepackage{multirow}
\usepackage{adjustbox}
\usepackage{booktabs}
\usepackage{braket}
\usepackage{xcolor}
\usepackage{xspace}

\newcommand{\Planck}{\textit{Planck}\xspace}
\newcommand{\thethree}{{\sc The300}\xspace}

\DeclareSIUnit{\parsec}{pc}
\DeclareSIUnit{\arcminute}{arcmin}
\DeclareSIUnit{\arcmin}{arcmin}
\DeclareSIUnit{\year}{yr}
\newcommand{\msun}{\ensuremath{\mathrm{M}_{\odot}}}
\newenvironment{enumerate*}{\begin{enumerate}%
        \setlength{\itemsep}{0pt}%
        \setlength{\parskip}{0pt}%
    }{\end{enumerate}}

\newcommand{\verify}[1]{{\color{orange} #1}}
\newcommand{\todo}[1]{{\bf \color{red} #1}}

\newcommand{\invisible}[1]{}

\newcommand{\const}[3]{\newcommand{#1}[1][]{{\verify{\ensuremath{#2\ifthenelse{\isempty{##1}}{\,#3}{}}}}}}

\arxivnumber{1234.56789} 
\title{The Atacama Cosmology Telescope: a census of bridges between galaxy clusters}

\author[1,2,3,4]{G.~Isopi,}
\author[1]{V.~Capalbo,}
\author[5,6]{A.~D.~Hincks,}
\author[7]{L.~Di~Mascolo,}
\author[1,2,4]{E.~Barbavara,}
\author[1,3,8]{E.~S.~Battistelli,}
\author[9]{J.~R.~Bond,}
\author[29]{W.~Cui,} 
\author[10,11]{W.~R.~Coulton,}
\author[1]{M.~De~Petris,}
\author[12]{M.~Devlin,}
\author[13]{K.~Dolag,}
\author[14,12]{J.~Dunkley,}
\author[15,16,17]{D.~Fabjan,}
\author[1]{A.~Ferragamo,}
\author[18]{A.~S.~Gill,}
\author[5]{Y.~Guan,}
\author[19]{M.~Halpern,}
\author[20,21]{M.~Hilton,}
\author[22]{J.~P.~Hughes,}
\author[23]{M.~Lokken,}
\author[24]{J.~van~Marrewijk,}
\author[25,26]{K.~Moodley,}
\author[23]{T.~Mroczkowski,}
\author[27]{J.~Orlowski-Scherer,}
\author[15,16,28]{E.~Rasia,} 
\author[1,29]{S.~Santoni,}
\author[26]{C.~Sifón,}
\author[30]{E.~J.~Wollack}
\author[31]{and G.~Yepes}

\affiliation[1]{Sapienza Università di Roma, Piazzale Aldo Moro, 5, 00185, Rome (RM), Italy}
\affiliation[2]{Università degli Studi di Roma Tor Vergata, Via Cracovia, 50, 00133, Rome (RM), Italy}
\affiliation[3]{INFN Sezione Roma1, Piazzale Aldo Moro, 2, 00185, Rome (RM), Italy}
\affiliation[4]{INAF OAC, Via della Scienza, 5, 09047, Selargius (CA), Italy}
\affiliation[5]{David A. Dunlap Department of Astronomy \& Astrophysics, University of Toronto, 50 St. George St., Toronto ON M5S 3H4, Canada}
\affiliation[6]{Specola Vaticana (Vatican Observatory), V-00120 Vatican City State}
\affiliation[7]{Laboratoire Lagrange, Université Côte d’Azur, Observatoire de la Côte d’Azur, CNRS, Blvd de l’Observatoire, CS 34229, 06304 Nice cedex 4, France}
\affiliation[8]{INAF IAPS, Via del Fosso del Cavaliere, 100, 00133 Roma RM, Italy}
\affiliation[9]{Canadian Institute for Theoretical Astrophysics, 60 St. George Street, University of Toronto, Toronto, ON, M5S 3H8, Canada}
\affiliation[10]{Kavli Institute for Cosmology Cambridge, Madingley Road, Cambridge CB3 0HA, UK}
\affiliation[11]{DAMTP, Centre for Mathematical Sciences, University of Cambridge, Wilberforce Road, Cambridge CB3 OWA, UK}
\affiliation[12]{Department of Astrophysical Sciences, Peyton Hall, Princeton University, Princeton, NJ 08544, USA}
\affiliation[13]{Ludwig-Maximilians-Universität München, Geschwister-Scholl-Platz 1,  D-80539 München, Germany}
\affiliation[14]{Joseph Henry Laboratories of Physics, Jadwin Hall, Princeton University, Princeton, NJ 08544, USA}
\affiliation[15]{INAF - Osservatorio Astronomico di Trieste, via Tiepolo 11, I-34143 Trieste, Italy}
\affiliation[16]{IFPU - Institute for Fundamental Physics of the Universe, Via Beirut 2, I-34014 Trieste, Italy}
\affiliation[17]{Faculty of Mathematics and Physics, University of Ljubljana, Jadranska cesta 19, 1000 Ljubljana, Slovenia}
\affiliation[18]{Department of Aeronautics and Astronautics, Massachusetts Institute of Technology, 77 Massachusetts Avenue, Cambridge, MA 02139, USA}
\affiliation[19]{Department of Physics and Astronomy, UBC, 6224 Agricultural Road, Vancouver, Canada}
\affiliation[20]{Astrophysics Research Centre, University of KwaZulu-Natal, Westville Campus, Durban 4041, South Africa}
\affiliation[21]{Wits Centre for Astrophysics, School of Physics, University of the Witwatersrand, Private Bag 3, 2050, Johannesburg, South Africa}
\affiliation[22]{Department of Physics and Astronomy, Rutgers University, 136 Frelinghuysen Road, Piscataway, NJ 08854, USA }
\affiliation[23]{Institut de Física d'Altes Energies (IFAE), The Barcelona Institute of Science and Technology, Campus UAB, 08193 Bellaterra (Barcelona), Spain}
\affiliation[24]{European Southern Observatory, Karl-Schwarzschild-Str. 2, Garching 85748, Germany}
\affiliation[25]{School of Mathematics, Statistics \& Computer Science, University of KwaZulu-Natal, Westville Campus, Durban 4041, South Africa}
\affiliation[26]{Instituto de Física, Pontificia Universidad Católica de Valparaíso, Casilla 4059, Valparaíso, Chile}
\affiliation[27]{Department of Physics and Astronomy, University of Pennsylvania, 209 South 33rd Street, Philadelphia, PA, 19104, USA}
\affiliation[28]{Department of Physics; University of Michigan, 450 Church St, Ann Arbor, MI 48109, USA}
\affiliation[29]{Departamento de Física Teórica, Facultad de Ciencias, Universidad Autónoma de Madrid, Modulo 8, E-28049 Madrid, Spain}
\affiliation[30]{NASA Goddard Spaceflight Center, 8800 Greenbelt Rd, Greenbelt, MD 20771, USA}
\affiliation[31]{Departamento de Física Teórica and  CIAFF, Universidad Autónoma de Madrid, Cantoblanco, 28049 Madrid, Spain}

\emailAdd{giovanni.isopi@roma1.infn.it}

\abstract{
According to cosmic microwave background (CMB) measurements, baryonic matter constitutes about $5\%$ of the mass-energy density of the universe. A significant population of these baryons, for a long time referred to as ``missing'', resides in a low density, warm-hot intergalactic medium (WHIM) outside galaxy clusters, tracing the ``cosmic web'', a network of large scale dark matter filaments.
Various studies have detected this inter-cluster gas, both by stacking and by observing individual filaments in compact, massive systems. 
In this paper, we study short filaments ($<\SI{10}{Mpc})$ connecting massive clusters ($\langle M_{500}\rangle\approx 3\times 10^{14}~\msun$) detected by the Atacama Cosmology Telescope (ACT) using the scattering of CMB light off the ionised gas, a phenomenon known as the thermal Sunyaev-Zeldovich (tSZ) effect. The first part of this work is a search for suitable candidates for high resolution follow-up tSZ observations. We identify four cluster pairs with an intercluster signal above the noise floor (S/N $>$ 2), including two with a tentative $>2\sigma$ statistical significance for an intercluster bridge from the ACT data alone.  
In the second part of this work, starting from the same cluster sample, we directly stack on ${\sim}100$ cluster pairs and observe an excess SZ signal between the stacked clusters of $y=(7.2^{+2.3}_{-2.5})\times 10^{-7}$ with a significance of $3.3\sigma$. It is the first tSZ measurement of hot gas between clusters in this range of masses at moderate redshift ($\langle z\rangle\approx 0.5$). We compare this to the signal from simulated cluster pairs with similar redshifts and separations in the THE300 and MAGNETICUM Pathfinder cosmological simulations and find broad consistency. Additionally, we show that our measurement is consistent with scaling relations between filament parameters and mass of the embedded halos identified in simulations. 
}

\begin{document}
\maketitle
\flushbottom

\section{Introduction}
\label{sec:intro}
In the standard $\Lambda$CDM model of contemporary cosmology, the dominant component of the universe is dark energy, which represents about $70\%$ of the total mass-energy density \citep{planck2018_VI,Hinshaw2013}. The remaining $30\%$ is the matter, dominated by cold dark matter (CDM), believed to be a component that only interacts through gravity. Baryonic matter comprises less than $5\%$ of the cosmic budget. Starting in the late 1990's, censuses of the baryon content of the universe found that the sum of all the contributions observed at the time accounted for only ${\sim}10\%$ of the ``cosmic baryon budget'' \cite{Fukugita1998} (see also \cite{Bregman_2007}). The search for missing baryons considered both the stellar and the non-stellar components of galaxy clusters, as X-ray observations revealed that their hot atmospheres, the intra-cluster medium (ICM), are several times more massive than the visible component \cite{physics_of_galaxy_clusters} (i.e. observable in the visible domain of the electromagnetic spectrum). Including the ICM mass (see, e.g., references \cite{WHIM_virgo_2023}, \cite{WHIM_coronaBorealis_2005} and \cite{WHIM_shapley_1998}) increases the observed baryon fraction to $12\%$. In addition, observations of neutral hydrogen through Ly-$\alpha$ forests at low redshift ($z<0.07$) \cite{Penton_2004,Sembach_2004} brought the content up to $~40\%$. 
However, a significant portion (${\sim}60\%$) of baryons was still waiting to be firmly detected. A broadly accepted solution to the missing baryon problem was proposed by Fukugita et al.~\cite{Fukugita1998}, who estimated that about one third of the baryons could be located in warm gas outside galaxy clusters, called the warm--hot intergalactic medium (WHIM). Hydrodynamic simulations \cite{Cen_1999,Tuominen_2021} suggest that these clouds trace the so-called cosmic web, the network of dark matter filaments connecting clusters. Cosmic filaments are expected to have low overdensities ($\delta \sim 10{-}100$) and to be collapsed but not virialized.

Recent work using the dispersion of fast radio bursts has shown that the baryon fraction integrated over the line of sight is indeed compatible with early universe measurements \citep{macquart/etal:2020,khrykin/etal:2024}. Such evidence indicates that baryons are not missing, but rather are hosted in low-density environments, which makes direct observations of their X-ray or millimetric signal challenging.

The existence of the cosmic web and the WHIM is widely accepted and supported by a number of observations, but the direct observation of individual filaments is challenging. Because of their low densities and relatively low temperature, cosmic filaments are very faint in X-rays (see, e.g., \cite{eROSITA_LOFAR_filaments}). Furthermore, the weakness of large-scale magnetic fields and the low density of relativistic electrons means that synchrotron radiation is faint (see, e.g., \cite{eROSITA_LOFAR_filaments, Vernstrom23_filaments_radio}). However, instead of detecting emission from these filaments, one can exploit the thermal component of the Sunyaev-Zeldovich (tSZ) effect \cite{SZ1969,SZ_1972} (for a review, see \cite{Mroczkowski_2019}).
The tSZ is a spectral distortion of the cosmic microwave background (CMB) produced by the interaction of low energy CMB photons with hot electrons permeating the ICM.
Its amplitude at a given frequency is proportional to the total pressure of the gas integrated along the line of sight and can be quantified by the Compton-$y$ parameter,
\begin{equation}
\label{eqn:tSZ}
    y = \frac{\sigma_{T}}{m_e c^2} \int k_B n_e(\ell) T_e(\ell) \, d\ell,
\end{equation}
where $\sigma_T$, $m_e$ and $c$ are the Thomson scattering cross-section, the electron mass at rest and the speed of light, respectively, $k_B$ is the Boltzmann constant, $n_e$ and $T_e$ are the electron density and temperature, respectively, and $\ell$ is the distance along the line of sight.
Due to its linear dependence on the temperature and electron density, in its hot phases it is more easily detected at low densities compared to the bremsstrahlung X-ray emission which, in turn, depends on the electron density squared.

To date, several observations with different radio and CMB experiments have shown evidence for the existence of filaments between clusters. The first detections were direct observations of single filaments between massive, local clusters. 
The \Planck Collaboration~\cite{Planck13_filaments} used a Compton-$y$ map at a resolution of $7.18'$ constructed from \Planck data to report the first direct tSZ detection of an intercluster filament between galaxy clusters. The authors investigated 25 cluster pairs, but only two systems, namely Abell~399 -- Abell~401 (A399–401) and Abell~3391 -- Abell~3395 (A3391--3395), exhibited a significant intercluster excess signal after accounting for the cluster contributions, with a projected length of respectively $\sim \SI{3}{Mpc}$ for both systems. 

A399--A401 was studied in detail in later works. Bonjean et al.~\cite{Bonjean2018} used \Planck data again to characterize the intercluster component. In addition, they examined the nature of galaxies within both the clusters and the intercluster region, finding that most are passive and red galaxies, with suppressed star formation, supporting the hypothesis that the entire system formed simultaneously and that the intercluster bridge could be a remnant of a cosmic filament heated during a pre-merging phase. These results were later confirmed by Hincks et al.~\citep{Hincks_2022}, who used Atacama Cosmology Telescope (ACT, see section~\ref{sec:ActFilaments}) and MUSTANG-2 to provide higher resolution and conduct a more precise investigation of this system.  In this work, they compared results from different physical models used to fit the two clusters and a bridge component, reporting evidence for a structure between the clusters at ${\approx}5.5\sigma$. For the bridge, they estimated a mean gas density of $n_e\sim 0.88 \times 10^{-4} \unit{cm^{-3}}$, significantly lower than previous estimates \cite{Planck13_filaments,Bonjean2018}. They explained these results by analysing the geometry of the system, concluding that we are viewing the  intercluster bridge at an acute angle because the two clusters are separated much more along the line of sight than in the plane of the sky. The same bridge was detected in the radio domain by Govoni et al.\ \citep{Govoni2019_lofar_TheBridge}, and these data enabled further constrains on the physics of the systems, by cross correlating with SZ and X-ray data. Indeed, further analysis of this pair was carried out by Radiconi et al.\ \citep{Radiconi2022} who, for the first time at high spatial resolution ($\sim$100\,kpc), studied the correlation between thermal and non-thermal emission of the system by combining X-ray, tSZ effect, and radio observations.

The second significant system from the \Planck Collaboration~\cite{Planck13_filaments}, A3391--A3395, is more difficult to analyse than A399--401 with the low angular resolution data from \Planck due to its complex morphology and dynamical state. They reported evidence for a residual signal in the intercluster region after subtracting the cluster components. However, the authors pointed out that the residual is suspiciously close to A3395, and that high angular resolution data would be needed to correctly model the system. A deep observation from eROSITA provided an unprecedented view of the complexity of the system, identifying multiple filaments and WHIM populations \citep{Reiprich2021, Veronica2024}. A high angular resolution investigation of the tSZ signal from this cluster pair is being conducted using ACT observations (Capalbo et al., in prep.).

The search for the faint tSZ signature of WHIM in cosmic filaments requires high sensitivity, which is often hard to achieve on a single target. Another approach is to use stacking, which consists in co-adding a large number of Compton-$y$ maps to increase the signal-to-noise ratio. In De Graaff et al.~\cite{de_Graaff_2019}, \Planck maps were aligned according to a catalogue of approximately one million galaxy pairs, built by selecting CMASS galaxies with small enough comoving ($5-14\,h^{-1}\,\si{\mega\parsec}$) and line-of-sight separation ($<5\,h^{-1}\,\si{\mega\parsec}$). A control sample was built by cutting out $13\times 10^6$ non-physical galaxy pairs, i.e., galaxy pairs with a large line of sight separation ($40-200h^{-1}$ Mpc), and thus unlikely to be connected by filaments. The resulting stack of the selected physically bound systems provided a $2.9\sigma$ evidence of an excess emission between the stacked galaxies, with an average amplitude of $\bar{y}=(0.6\pm 0.2)\times 10^{-8}$. This translates to a baryon fraction of $11\pm7\%$, with a large uncertainty due to the assumed electron temperature provided by simulations. This result is consistent with Tanimura et al.\ \cite{tanimura_2019}, in which a similar analysis is presented, but using a catalogue of $2.6\times 10^5$ luminous red galaxies (LRGs) that are typically the central galaxies in massive halos ($M \approx 10^{13}~{\msun}$). The amplitude of the excess signal between the clusters is $\bar{y}\approx 1.31 \times 10^{-8}$ with a significance of $3.5\sigma$, consistent with De Graaff et al.\ \cite{de_Graaff_2019}, with an overdensity of $\delta_c\approx 3$. A more recent study by Singari et al.\ \cite{Singari_2020} repeated the analysis using a sample of $8.8 \times 10^4$ LRG pairs, with an enhanced selection criterion to minimize the galactic contamination and a different approach, by first stacking the single-frequency maps and then applying the component separation. The amplitude of the excess is $\bar{y}\approx 3.8 \times 10^{-8}$, larger than Tanimura et al.\ \cite{tanimura_2019}, with an overdensity of $\approx 10$, but with a higher significance of $10.2\sigma$ due to the more robust foreground removal.

While these past works provided first detections of filamentary gas in individual systems and early studies of intercluster and intergalactic filamentary gas using the tSZ, we now aim to leverage the high sensitivity and increased resolution of ACT data to explore a sample of high-mass galaxy cluster pairs. 

This paper is structured as follows. In section~\ref{sec:ActFilaments} we introduce the cluster catalogue derived from ACT data used in this study. In section \ref{sec:phys_models} we introduce the models and the fitting method. In section~\ref{sec:single_pairs} we search for filaments in individual pairs of galaxy clusters. In section~\ref{sec:stack} we stack the cluster pairs in the catalogue to study their average properties with an approach similar to past works with \Planck data \citep{de_Graaff_2019, tanimura_2019, Singari_2020}. 
In section~\ref{sec:sims} we compare the results of the stack with THE300 and MAGNETICUM-Pathfinder simulations, and in section~\ref{sec:discussion} and \ref{Sect:Conclusions} we discuss our results and conclude. All the results and calculations assume a $\Lambda$CDM flat universe model, with cosmological parameters from Planck18 \cite{planck2018_VI}. Unless otherwise noted, uncertainties are quoted at $1\sigma$.


\section{Data}
\label{sec:ActFilaments}
ACT\footnote{\url{https://act.princeton.edu/}} was a \SI{6}{m} off-axis Gregorian telescope that operated from 2007 to 2022 from the Atacama Desert in Chile at an altitude of 5190\,m. It observed the CMB with three generations of receivers: MBAC, ACTPol and AdvACT. The MBAC receiver used ${\sim}3000$ transition edge sensor (TES) detectors to observe in bands centered at 148, 218 and 277\,GHz without polarization sensitivity \cite{ACT_overview}. ACTPol was the first polarization sensitive receiver to be installed on the ACT platform, and used two arrays of TES at 146\,GHz and a third array with dichroic TES observing at 146 and 90\,GHz \cite{ACTPol_overview}. The final iteration, AdvACT, initially comprised 3072 TES detectors operating in 3 bands centred at 98, 150, and 220\,GHz \cite{AdvACT_detectors}. Two additional bands, centred at 27 and 39\,GHz, were added in 2020 \cite{Koopman_2018_ACT_LF}.

ACT maps and derived products are available on the LAMBDA website in a series of data releases (DR1--6),\footnote{\url{https://lambda.gsfc.nasa.gov/product/act/actpol_prod_table.html}} the most recent complete release being DR6 which included data from 2008 to 2022.
Hilton et al.\ 2021~\cite{Hilton_2021} presented a catalogue of more than 4000 SZ-selected clusters in DR5. The catalogue was built from the $98$ and $\SI{150}{GHz}$ maps, which are at frequencies where the tSZ spectral distortion is negative and near where its amplitude peaks. In these bands, ACT had a resolution of $2'$ and $1.4'$, respectively, on more than $\SI{18000}{deg^2}$ of the sky, with varying noise level, although only $\SI{12000}{deg^2}$ were used for the blind search because of contamination from dust emission as well as high density of stars at low Galactic latitudes, that could make the optical confirmation more difficult \cite{Hilton_2021}. Clusters were blindly detected using a multi-frequency matched filter, using a Universal Pressure Profile \cite{arnaud2010} with a minimum signal-to-noise-ratio of 4. Among these clusters, about $140$ pairs of likely physically bound clusters were flagged. The DR5 tSZ cluster catalogues, including a dedicated catalogue of cluster pairs, are available on LAMBDA.\footnote{\url{https://lambda.gsfc.nasa.gov/product/act/actpol_dr5_szcluster_catalog_get.html}}

\begin{figure}
    \centering
    \includegraphics[width=\textwidth]{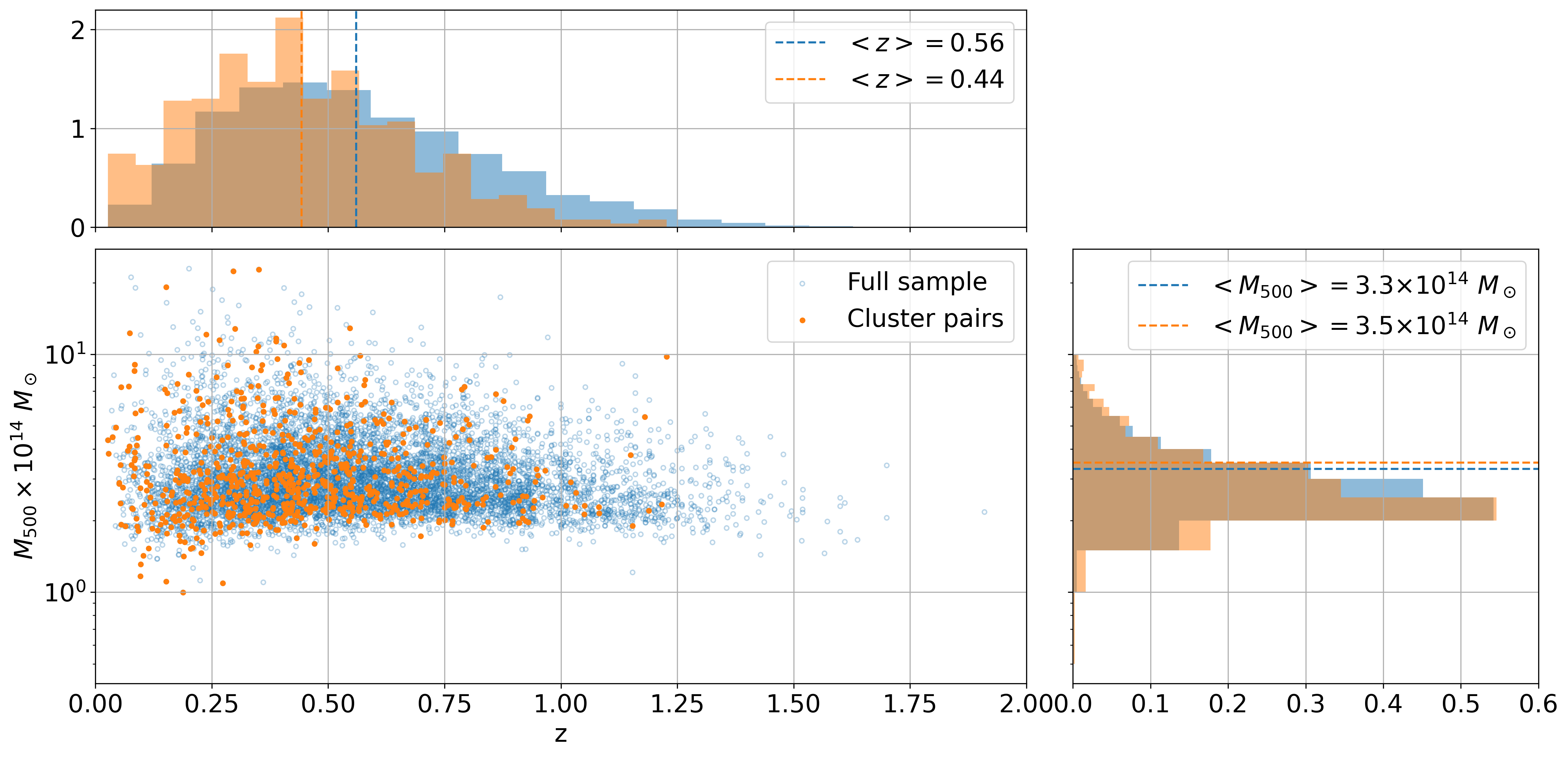}
    \caption{Scatter plot of the cluster sample used in this work, plotted on the $z-M_{500}$ plane. Both redshift and mass are provided by the cluster catalog. The empty dots show the distribution of the single clusters in the catalog. Orange filled dots show the subsample of individual clusters that belong to a candidate pair in the double cluster catalog.}
    \label{fig:mass-z-scatter}
\end{figure}

The Compton-$y$ map used in this paper is described in Coulton et al. 2024 \cite{Coulton24_ymap}, which includes data from DR4 and DR6, collected from 2013 to 2022. A catalogue of galaxy clusters based on DR6 is in preparation; in this paper we use a preliminary version of this catalogue containing $\approx7400$ systems (ACT Collaboration, in prep.; this number should considered as preliminary). We have identified pairs of clusters in the catalogue that are physically close as determined by their projected distance on the plane of the sky, $d_{\mathrm{sky}}<10~\unit{Mpc}$, and their peculiar velocity difference, $\Delta v_z<5000~\unit{km~s^{-1}}$, as was done for the DR5 cluster catalogue \citep{Hilton_2021}. A total of 284 systems meet these criteria, including clusters with only photometric redshift. They fall in the redshift range $0.04<z_{\mathrm{pair}}<1.25$, where $z_{\mathrm{pair}}$ is the mean redshift between the two clusters of each pair. Figure~\ref{fig:mass-z-scatter} shows the $z_{\mathrm{pair}}$ distribution and the $M_{500}$ distribution for clusters in pairs and for the full sample. The mean calibrated mass for clusters in multiple systems is $3.5\times 10^{14}\SI{}{\msun}$, slightly larger than the mean mass of the full sample, which is $3.3\times10^{14}\msun$. For the stacking analysis in this paper (section~\ref{sec:stack}), however, we only use cluster pairs which have spectroscopic redshifts available, because the uncertainty of the photometric redshift, when propagated to $\Delta v_{z}$, is comparable with the $\SI{5000}{km/s}$ threshold used to select candidate bound systems. This translates into a large fraction of false positives in the sample, reducing the intercluster signal by about a factor of two when they are included in the stack (see section~\ref{sec:null_test}). Moreover, we exclude from our analysis systems containing more than two components. After this selection, the number of stacked cluster pairs is 86. The $5,50,95th$ quantiles of this sample are respectively $[1.9,3.1, 6.7]\times 10^{14}~\msun$. The lack of spectroscopic redshift measurement is not correlated to cluster properties in our sample, but it is due to the lack of coverage by spectroscopic surveys in the southernmost parts of the ACT footprint. However, we do not exclude photometric systems from our search of filaments in individual pairs, since they might still host a significant filament signal. Note that our catalogue does not include A3395--A3391 or A399--A401, the two pairs hosting observable filaments described above (section~\ref{sec:intro}), since their low redshift makes their angular size too large to be detected efficiently by the matched filter used to construct the cluster catalogue, even if their tSZ signal is strong.
In this work, we do not make additional effort to include these known, low-redshift clusters to avoid being contaminated by systems that could include outliers with exceptionally bright bridges.
We are preparing a separate study on the noteworthy A3395-3391 (Capalbo et al., in prep) which has not yet been examined with ACT Compton-$y$ maps.
We use the most recent ACT Compton-$y$ map to analyze the intercluster tSZ signal in our catalogue of galaxies cluster pairs. Details of this map are provided in Coulton et al.\ 2024~\cite{Coulton24_ymap}, but here we provide a brief summary. The $y$-map is constructed from ACT DR4 and DR6 maps at approximate frequencies of 93, 148 and 225\,GHz, as well as  the \Planck NPIPE maps \cite{Planck_NPIPE} from 30 to 857\,GHz. The \Planck maps are included to provide sensitivity at larger angular scales where ACT maps become noisy due to the emission of atmospheric water vapour, and to add additional frequency channels for a more robust foreground removal. The multiple frequencies are combined via an internal linear combination (ILC) which exploits the differences in the spectra of the tSZ signal, the CMB blackbody signal and, optionally, thermal emission from dust to better isolate the tSZ component at the expense of a higher noise level. The authors used needlet basis functions to better account for variations in signal and noise across different regions of the input maps than previous approaches that worked in harmonic space. The resulting $y$-map has a resolution of $1.6'$. As mentioned above, versions of the map that deproject a dust-like spectrum are available, but we do not use these because of their higher noise. We however checked if there was a significant difference between the results obtained with the two maps, and found no significant difference. For details on this, see appendix~\ref{apx:cib}. There are also variants of the map that use different masks of the Galactic regions; in this paper, we use the \texttt{mask80} map, which gives a large footprint to include as many cluster pairs as possible.

\section{Models}

\label{sec:phys_models}
In order to reveal the signature of a bridge of matter between the clusters, whether in individual pairs (section~\ref{sec:single_pairs}) or in stacks (section~\ref{sec:stack}), we need to model the different components in the $y$-maps, i.e., the clusters and the bridge region.

For the clusters we adopt the generalized Navarro, Frenk \& White (gNFW) pressure profile \citep{Nagai_2007,arnaud2010} given by:
\begin{equation}
    \label{eq:gnfw_circ}
    P(r) = \frac{P_0}{\Big( \frac{r}{r_s} \Big)^\gamma \Big[ 1+ \Big(\frac{r}{r_s} \Big)^{\alpha} \Big]^{(\beta-\gamma)/\alpha}},
\end{equation}
where $r$ is the radial distance from the cluster's centre, $P_0$ is the central pressure, $r_s=R_{500}/c_{500}$, with $c_{500}$ being the concentration parameter, and where $\gamma$, $\alpha$ and $\beta$ are the central ($r \ll r_s$), intermediate ($r \sim r_s$) and outer ($r \gg r_s$) slopes of the profile, respectively.
The gNFW profile is spherically symmetric and cannot account for possible asphericity in the cluster's geometry. Therefore, we also use an ellipsoidal extension to the profile \citep{DiMascolo2019}. It is given by replacing the ratio $(r/r_s)$ in eq.~\ref{eq:gnfw_circ} with
\begin{equation}
    \label{eq:gnfw_ell}
    \xi = \frac{1}{r_s} \Big( r_a^2 + \frac{1}{1-\epsilon^2}r_b^2 + \frac{1}{2}\frac{2-\epsilon^2}{1-\epsilon^2}r_c^2 \Big)^{1/2}
\end{equation}
where $r_a$, $r_b$ and $r_c$ are the radii along the three principal axes of the ellipsoid, respectively, and $\epsilon$ is the eccentricity of the profile projected on the sky. Let $\theta$ be the position angle of the plane-of-sky major axis; then the equations for $r_a$, $r_b$ and $r_c$ are:
\begin{equation}
    \begin{split}
        & r_a = (x-x_0)\cos y_0 \cos\theta - (y-y_0)\sin\theta, \\
        & r_b = (x-x_0)\cos y_0 \sin\theta + (y-y_0)\cos\theta, \\
        & r_c = z-z_0,
    \end{split}
\end{equation}
where $(x-x_0)$, $(y-y_0)$ and $(z-z_0)$ are differences, on the three axes, of a given point with respect to the centre of the ellipsoid, $(x_0,y_0,z_0)$. For simplicity, in the fit we assume that two of the three axes (clearly corresponding to x and y) are on the plane of the sky, while the other one is aligned along the line of sight.

To model the bridge region we use an isothermal $\beta$-model \citep{beta_model} with cylindrical geometry \citep{arnaud2010,Bonjean2018}. In this case, the pressure profile is given by
\begin{equation}
    \label{eq:beta_cyl}
    P(r)=\frac{P_0}{\Big[ 1 + \Big( \frac{r}{r_c} \Big)^2 \Big]^{\frac{3}{2}\beta}},
\end{equation}
where $P_0$ is the central pressure, $r_c$ is the core radius and $\beta$ is the slope fixed to $4/3$, as suggested by Ostriker et al.\ \citep{Ostriker1964} for a non-magnetised filament. The model is calculated over a length $l_{fil}$. 

We follow the formalism of Yoshikawa et al.\ \citep{Yoshikawa1999} \citep[see also][]{arnaud2010} to integrate the pressure profiles described above along the line of sight, i.e., to project them from 3D to 2D in the plane of the sky. All the models shown above are written in terms of the central pressure $P_0$. Since our maps are in terms of Compton-$y$ parameter, which is proportional to $P_0$, the amplitude $A_{cl}$ we fit for clusters and $A_{fil}$ for filaments is a Compton-$y$ amplitude. 

\subsection{MCMC fit}
\label{sec:mcmc_fit}
We perform a Markov chain Monte Carlo (MCMC) analysis in order to constrain the best values for the parameters in the models described in section~\ref{sec:phys_models}.
We use the {\sc emcee} \citep{emcee} Python package to estimate the posterior distribution for the parameters, using the following likelihood:
\begin{equation}
    \label{eq:likelihood}
    \mathcal{L}=\frac{1}{2\pi|\mathbf{M}|^{1/2}} \exp \left( -\frac{1}{2} \mathbf{m}^T \mathbf{M}^{-1} \mathbf{m} \right),
\end{equation}
where $\mathbf{m}$ is the difference between the $y$-map and the model map and $\mathbf{M}$ is the covariance matrix.
Here, we follow the same approach as in Hincks et al.\ \citep{Hincks_2022}, computing the likelihood in the Fourier space in which the noise covariance is diagonal under our assumption of stationary noise. We estimate $\mathbf{m}$ after convolving the model with a Gaussian kernel of $1.6^{\prime}$ FWHM to match the angular resolution of the $y$-map. To compute the covariance matrix we use a maximum of 8 external fields around the $y$-map that is being fit, placed in an analogous way as in Hincks et al. (2022) \cite{Hincks_2022}. Only fields with sufficiently Gaussian noise are used, which we determine with the Kolmogorov-Smirnov (KS) test; unwanted non-Gaussianity in the $y$-signal used to estimate the noise covariance can come, for example, from background sources and foreground contamination. The Gaussian noise assumption is an approximation, but for our relatively small maps is reasonable, as for A399--A401 \cite{Hincks_2022}. We repeat this selection of fields for both single pairs and the stacked map. In the stacked map, the averaging of the noise from different sky patches helps to average out any non-Gaussianity, further supporting our assumption.
We evaluate convergence in a similar manner to Hincks et al. (2022) \cite{Hincks_2022}. Namely, we estimate the autocorrelation length of each parameter $\tau_i$ every 100 iterations and define $\tau = \max(\tau_i)$. We run the chains until the number of samples $N > 20\tau$; that is we assume that the estimate of $\tau$ is reasonably accurate when the chains are 20 times longer than this estimate. Then, we ensure that none of $\tau_i$ increases more than 1\% within 100 iterations before stopping the MCMC. Finally, we treat the first $N_{\mathrm{burn}}=10\tau$ iterations as burn-in, i.e., we calculate the posteriors after rejecting the first $N_{\mathrm{burn}}$ samples. Hincks et al. (2022) \cite{Hincks_2022} showed that for the Abell 399--401 bridge system, which was fit with a similar model, that this was a reliable test of convergence. In our case the autocorrelation length of the bridge length parameter, $l_{\mathrm{fil}}$ is still slowly increasing when we reach this point, which we attribute to the double-peaked posterior. However, because $l_{\mathrm{fil}}$ is not correlated with the filament amplitude, $A_{\mathrm{fil}}$, the main parameter of interest, we do not consider this worrisome. Nevertheless, we ran an additional MCMC with chain lengths \~25 times longer than the one used for our results on the stacked map using the 2GNFW+cyl-$\beta$ model to test whether $l_{\mathrm{fil}}$'s autocorrelation length diverges in longer chains, We found that all parameters, included $l_{\mathrm{fil}}$, still satisfy the convergence criterion that we adopt.

We perform four different fits on each $y$-map that refer to different combinations of the physical models in the previous sections:
\begin{itemize}
    \item `2gNFW$_{\mathrm{sph}}$' -- The simplest fit includes two gNFW spherical pressure profiles (eq.~\ref{eq:gnfw_circ}) for the clusters. It has four free parameters: the external slopes $\beta_{1,2}$ (where the subscripts ``1'' and ``2'' refer to each of the clusters) and the central amplitudes $A_{1,2}$.  We fix the coordinates of the clusters, the scale radius $r_s=R_{500}/c_{500}$, where $c_{500}=1.81$, and we fix the inner and mid slopes $\gamma = 0.31$ and $\alpha = 1.33$, as well as $c_{500}$, according to the average profiles measured by \Planck  \citep{PlanckClusterProfiles2013}. 
    \item `2gNFW$_{\mathrm{sph}}$+cyl-$\beta$' -- The second fit extends `2gNFW$_{\mathrm{sph}}$' by including a bridge component described by the cylindrical $\beta$-model in eq.~\ref{eq:beta_cyl}. It has six free parameters: the four in `2gNFW$_{\mathrm{sph}}$', plus the cylinder's core radius $r_c$ (i.e. the width of the bridge) and its amplitude $A_{fil}$. The length is fixed to the distance between the two $R_{500}/2$ values of the two members of the cluster pairs and its central coordinates are the average of the coordinates of the two clusters.
    \item `2gNFW$_{\mathrm{ell}}$' -- This fit is an extension of `2gNFW$_{\mathrm{sph}}$' that allows the pressure profiles of the clusters to be elliptical (eq.~\ref{eq:gnfw_ell}). It has eight free parameters: the four from `2gNFW$_{\mathrm{sph}}$', plus an eccentricity, $\epsilon$, and an angle, $\theta$, for each of the two clusters.
    \item `2gNFW$_{\mathrm{ell}}$+cyl-$\beta$' --- The final fit extends `2gNFW$_{\mathrm{ell}}$' by adding a bridge component, analogously to `2gNFW$_{\mathrm{sph}}$+cyl-$\beta$'. It has ten free parameters.
\end{itemize}

We employ two statistical tests to compare the various models and determine the most suitable one for fitting the data. First, we compute the likelihood ratio between two models:
\begin{equation}
    W = 2 \log \frac{\max \mathcal{L}_2}{\max \mathcal{L}_1},
    \label{eq:likelihood_ratio}
\end{equation}
where $\max \mathcal{L}$ is the maximum likelihood computed with the best-fitting parameters. This test requires that Model 1, the simpler one, be nested in Model 2. Given the null hypothesis that Model 1 is true, for a large number of data points $W$ approaches a $\chi^2$ distribution with degrees of freedom equal to the difference in the number of parameters between the two models \cite[][\S5.5]{held/bove:2020}. We compute the $p$-value of the null hypothesis and the associated  $\sigma$ of the alternative. Note that this test is more rigorous than the S/N estimate (Equation \ref{eq:snr}) to select the pairs for fitting, and is not expected to give the same results.

The second statistical test is the Akaike Information Criterion (AIC) \citep{AIC}. This is a more general method that does not require that the models be nested. The test statistic is:
\begin{equation}
    \mathrm{AIC} = 2 k - 2 \log (\max\mathcal{L}),
    \label{eq:AIC}
\end{equation}
where $k$ is the number of free parameters in the model.
Models with smaller AIC values offer better descriptions of the data. However, AIC is a relative measure rather than an absolute one and the difference $\Delta$AIC between two models is the relevant quantity. Let AIC$_0$ be the smallest value, indicating the preferred model and let $\Delta_{\text{AIC},i}$ = AIC$_i$ - AIC$_0$ be the difference between the $i$-th model and the preferred one. According to Ref.~\citep{delta_AIC}, if $\Delta_{\text{AIC},i} \lesssim 2$, the two models provide roughly the same quality fit; if $\Delta_{\text{AIC},i} \gtrsim 4$, the model with the higher AIC value, AIC$_i$ has ``considerably less support'' than the model with AIC$_0$, and if $\Delta_{\text{AIC},i} \gtrsim 10$, it has ``essentially no support'' compared to the AIC$_0$ model.

\section{Analysis of Single Pairs of Clusters}
\label{sec:single_pairs}

In the first part of this work we analyze the cluster pairs individually in order to reveal potential bridges of diffuse matter associated with WHIM that may serve as targets for future follow-up observations at improved angular resolution using millimeter cameras coupled to large, single dish telescopes. In particular, we exploit the higher angular resolution of ACT to extend the analysis to higher redshift systems ($z\approx 0.5$) compared with previous work using {\Planck}-only data \citep{Planck13_filaments, Bonjean2018}, which could only resolve clusters in the local universe.

\subsection{Methods}
We select a subsample of systems from our catalogue of cluster pairs by estimating the significance of the tSZ signal in the intercluster region, following the approach used in Bonjean et al.\ \cite{Bonjean2018}. For each cluster pair, a cutout from the full $y$-map is created that is centred on the mean position of the cluster pair and has both dimensions equal to five times the projected separation of the clusters. The cutout is then reprojected onto a tangent plane (i.e. gnomonic reprojection). 
We exclude from our consideration any pairs in which the clusters are not sufficiently separated, namely, in which the distance between the cluster centres is smaller than the sum of their two $R_{500}$, as listed in the preliminary ACT DR6 catalogue.
We then define a rectangular region in the centre of the map (see figure~\ref{fig:single_pairs} for a visualization) with length equal to the distance between the two $R_{500}/2$ and width equal to the minimum of the two $R_{500}$, inside of which we compute the mean tSZ signal, $\hat{y}$. For the noise estimation, we consider the background area $3\times R_{500}$ beyond the cluster centres and mask possible background sources with a $3\sigma$-clip. The signal-to-noise estimation (henceforth S/N, to differentiate from the statistical evidence $\sigma$ introduced below) is given by:
\begin{equation}
    \label{eq:snr}
    S/N = \frac{\hat{y} - \hat{y}_{\mathrm{ext}}}{\sigma_{\mathrm{ext}}},
\end{equation}
where $\hat{y}_{\mathrm{ext}}$ and $\sigma_{\mathrm{ext}}$ are the mean tSZ signal and the standard deviation of the background area, respectively. Again following Bonjean et al.~\cite{Bonjean2018}, we select pairs with S/N~$\ge2$ for further investigation. The four systems that meet this criterion are listed in table~\ref{tab:single_pairs} and their $y$-maps are shown in figure~\ref{fig:single_pairs}. For convenience, the maps are rotated counterclockwise so that the axis connecting the two cluster centres lies along the $x$-axis.

We fit the 2D maps with the models described in section \ref{sec:mcmc_fit}. Because of the low signal to noise ratio of the single pairs and the limited resolution of the maps, we fixed different parameters depending on the case. In all the pairs, we fix the size of the bridge to the same box used for the $S/N$ estimation, and only leave $A_{fil}$ free. 

\begin{table*}

\begin{adjustbox}{width=1\textwidth}
     \centering
     \begin{tabular}{ccccccc}%
         \hline\hline
         Name & RA & Dec & $z$ & $R_{500}$ & $D_{sky}$\\
         & (deg) & (deg) &      & (arcmin) & (arcmin) \ (Mpc) & S/N\\
         \hline\hline
         ACT-CL~J2336.0$-$3210 (C$_1$)  & 354.016 & $-$32.176 & 0.613 & 2.39 & \multirow{2}{*}{5.55 \ \ \ \ \ 2.25} & \multirow{2}{*}{3.2}\\
         ACT-CL~J2336.3$-$3206 (C$_2$)  & 354.082 & $-$32.102 & 0.619 & 2.40 \\\\
         \hline
         ACT-CL~J0328.2$-$2140 (C$_1$)  & 52.054 & $-$21.668 & 0.590 & 2.78 & \multirow{2}{*}{5.23 \ \ \ \ \ 2.08} & \multirow{2}{*}{3.6} \\
         ACT-CL~J0328.5$-$2140 (C$_2$)  & 52.147 & $-$21.674 & 0.590 & 2.30 \\\\
         \hline
         ACT-CL~J0245.9$-$2029 (C$_1$)  & 41.489 & $-$20.486 & 0.310 & 3.57 & \multirow{2}{*}{7.78 \ \ \ \ \ 2.12} & \multirow{2}{*}{2.9}\\
         ACT-CL~J0246.4$-$2033 (C$_2$)  & 41.604 & $-$20.557 & 0.318 & 4.17 \\\\
         \hline 
         ACT-CL~J0826.0+0419 (C$_1$)  & 126.518 & 4.325 & 0.473 & 3.09 & \multirow{2}{*}{5.87 \ \ \ \ \ 2.08} & \multirow{2}{*}{2.3}\\
         ACT-CL~J0826.4+0416 (C$_2$)  & 126.600 & 4.271 & 0.463 & 2.33 \\\\
        \hline         
    \end{tabular}%
\end{adjustbox}
\caption{Physical properties of galaxy cluster pairs selected from ACT DR6 catalogue with potential gas bridges. $D_{sky}$ is the center-to-center projected distance on the sky.}
\label{tab:single_pairs}

\end{table*}

\subsection{Results and discussion}
\label{sec:results_single_pairs}
Among the 284 candidate interacting clusters, we found four pairs with S/N~$>2$. We fitted each of the four models described in section~\ref{sec:mcmc_fit} to these pairs and computed the statistics. The selected pairs are shown in table~\ref{tab:spstats}, while the best fit parameters are in table~\ref{tab:spfparams}. We leave a detailed description of each cluster system and the best fit parameters to appendix~\ref{apx:singlePairs}.
As mentioned at the beginning of this section, the goal of this analysis was to identify promising candidates for detecting intercluster gas signatures and for conducting high-resolution follow-up observations. It is clear how revealing individual filamentary structures between galaxy clusters is extremely challenging. From an initial sample of $>$ 200 systems it is possible to infer for only two pairs, namely J2336 and J0245, the presence of a bridge at low significance ($>2\sigma$ using the likelihood ratio $W$). Note the discrepancy between likelihood ratio and the raw S/N. It is important to emphasize, however, that these are  high redshift objects and are barely resolved by ACT. We note, interestingly, that in none of these systems the elliptical models are preferred with respect to the spherical models. This could be due to the low S/N of the maps. Additionally, though, an inter-cluster excess due to a bridge is not well modeled by only two elliptical pressure profiles because an enhancement of signal between in the inter-cluster region requires additional signal on the opposite side of each cluster, due to the symmetry of the elliptical models. If this signal outside the cluster pair is not present, the elliptical model is disfavoured; this was a finding of Hincks et al. (2022) \citep{Hincks_2022} on the high S/N A339--401 system .

Another possible explanation of the excess we observe is that we are not detecting a filament, but rather ICM in the outskirts of the clusters, that cause a deviation from the ideal gNFW profiles due to interaction between the clusters.

This underscores the necessity of obtaining higher angular resolution and higher sensitivity data for further study of these potential filaments in this redshift range, once they have been identified by wide area surveys by mid-resolution instruments like ACT. With this approach, high resolution cameras like MUSTANG-2 \citep{dickermustang2} or MISTRAL \citep{battistelliMistral} will enhance the detection of these systems in the same way as ACT enhanced the detections by \Planck. For this reason, we choose to report the selected systems, although it is not possible to firmly detect the candidate filaments with current data. 

\begin{table}
    \centering
    \begin{adjustbox}{width=1\textwidth}
    \begin{tabular}{ccccccc}%
    \hline\hline
     \multirow{2}{*}{Cluster Pair}\\ \multirow{2}{*}{[S/N]} & Model & \# free parameters & \multicolumn{3}{c}{Likelihood ratio} & $\Delta_{\text{AIC},i}$ \\
     \cmidrule(lr){4-6}
          &       &           & $W$ & $p$-value & $\sigma$           & \\
    \hline\hline
    \multirow{4}{*}{ACT-CL~J2336.0$-$3210}\\ \multirow{4}{*}{ACT-CL~J2336.3$-$3206} \\ \multirow{4}{*}{[3.2]}
            & 2gNFW$_{\mathrm{sph}}$             & 4  & --   & --                  & --   & 3.57 \\
            & 2gNFW$_{\mathrm{sph}}$+cyl-$\beta$ & 6  & 7.57 & $2.27\times10^{-2}$ & 2.28 & 0 \\
            & 2gNFW$_{\mathrm{ell}}$             & 8  & --   & --                  & --   & 12.31 \\
            & 2gNFW$_{\mathrm{ell}}$+cyl-$\beta$ & 10 & 9.09 & $1.06\times10^{-2}$ & 2.55 & 7.22 \\\\
    \hline
    \multirow{4}{*}{ACT-CL~J0328.2$-$2140}\\ \multirow{4}{*}{ACT-CL~J0328.5$-$2140}  \\ \multirow{4}{*}{[3.6]}
            & 2gNFW$_{\mathrm{sph}}$             & 4  & --  & --                   & --  & 0.01 \\
            & 2gNFW$_{\mathrm{sph}}$+cyl-$\beta$ & 5  & 2.01 & $1.56\times10^{-1}$ & 1.42 & 0 \\
            & 2gNFW$_{\mathrm{ell}}$             & 8  & --  & --                   & --  & 8.57 \\
            & 2gNFW$_{\mathrm{ell}}$+cyl-$\beta$ & 9  & 1.97 & $1.61\times10^{-1}$ & 1.40 & 8.60 \\\\
    \hline
    \multirow{4}{*}{ACT-CL~J0245.9$-$2029}\\ \multirow{4}{*}{ACT-CL~J0246.4$-$2033}  \\ \multirow{4}{*}{[2.9]}
            & 2gNFW$_{\mathrm{sph}}$             & 4  & --   & --                  & --  & 1.53 \\
            & 2gNFW$_{\mathrm{sph}}$+cyl-$\beta$ & 5  & 3.53 & $6.03\times10^{-2}$ & 1.88 & 0 \\
            & 2gNFW$_{\mathrm{ell}}$             & 8  & --   & --                  & --  & 5.14 \\
            & 2gNFW$_{\mathrm{ell}}$+cyl-$\beta$ & 9  & 4.98 & $2.57\times10^{-2}$ & 2.23 & 2.17 \\\\
    \hline
    \multirow{4}{*}{ACT-CL~J0826.0+0419}\\ \multirow{4}{*}{ACT-CL~J0826.4+0416}  \\ \multirow{4}{*}{[2.3]}
            & 2gNFW$_{\mathrm{sph}}$             & 4  & --   & --                  & --   & 0 \\
            & 2gNFW$_{\mathrm{sph}}$+cyl-$\beta$ & 5  & 0.33 & $5.69\times10^{-1}$ & 0.57 & 1.68 \\
            & 2gNFW$_{\mathrm{ell}}$             & 8  & --   & --                  & --   & 7.02 \\
            & 2gNFW$_{\mathrm{ell}}$+cyl-$\beta$ & 9  & 2.16 & $1.42\times10^{-1}$ & 1.47 & 6.86 \\\\ 
    \hline        
    \end{tabular}
    \end{adjustbox}
      \caption{Statistical tests for model comparison in the analysis of individual cluster pairs. The first column reports the name of the cluster pair and the preliminary S/N estimate of a bridge component, as described in section~\ref{sec:single_pairs}. In the second column there are the models adopted in the fits. The number of free parameters for each model is in the third column. $W$ is the likelihood ratio (4th column), $p$-value is the probability of the null hypothesis, i.e., the simpler model with less free parameters is true (5th column), and $\sigma$ is the significance of the rejection of the null hypothesis (6th column). $\Delta_{\text{AIC},i}$ (7th column) is the difference of AIC statistics of a model with respect to the preferred model, indicated with $\Delta_{\text{AIC},i}=0$. }
      \label{tab:spstats}

\end{table}

\section{Stack of cluster pairs}
\label{sec:stack}

The second part of this work focuses on stacking ACT selected cluster pairs.
\subsection{Methods}
\begin{figure}
    \centering
    \includegraphics[width=\columnwidth]{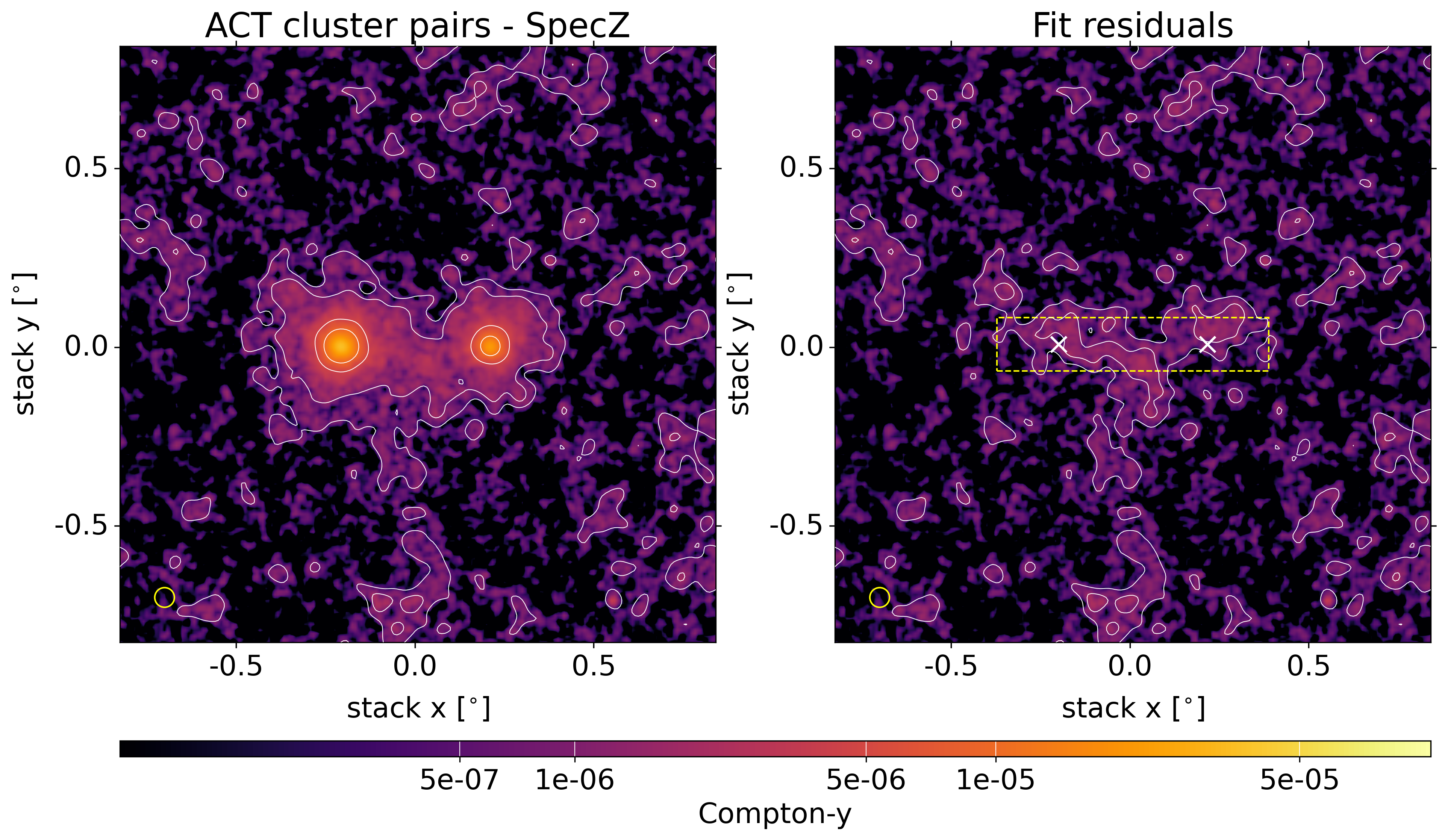}
    \caption{Stack of cluster pairs showing the candidate filament. Left: oriented stack of 86 clusters with spectroscopic redshift measurement for both clusters, with primary cluster on the left. Right: residual after subtracting the best fitting cluster profiles shown in section~\ref{sec:stack_mcmc}. A filamentary excess is visible between the two stacked clusters. Cluster positions in the right map are marked with white crosses. The yellow dashed box shows, for reference, the size $l_{\mathrm{fil}}\times 2\times r_{c,fil}$ of the best fitting cylindrical $\beta$-model. Note that $r_{c, fil}$ does not mark a precise boundary of the filament, therefore it is expected that some signal will fall outside of the box. Contour levels are $y=(0.5,1,5,10,50)\times 10^{-6}$, and they are smoothed with a \SI{3}{px} gaussian kernel for better visualization. The effective beam of $1.6\SI{}{'}$ is shown as a yellow circle in the bottom-left corner.}
    \label{fig:realStacks}
\end{figure}

We use a set of cut-outs extracted from the full Compton-$y$ map using the \texttt{pixell} map manipulation Python library\footnote{https://pixell.readthedocs.io/en/latest/} . For each cluster pair in the catalogue, we cut out a box centered on the average coordinates of the two components, with a radius equal to 10 times the clusters' angular separation. Each cut-out is then reprojected to a tangent-plane (or gnomonic) projection from the original \textit{Plate Carrée} projection and rotated such that the axis between the two clusters is aligned with the $x$-axis. We identify which cluster has the higher mass in the catalogue and flag it as the ``primary" cluster, while the other is the ``secondary" cluster. The rotation angle is chosen such that the primary cluster is placed on the left side of the image. We then scale the map using a bi-linear interpolation in order to place the two clusters at a fixed distance expressed in pixels. The value used in this work is 50 pixels, or $\SI{25}{'}$, because it is close to the average sky separation of the clusters and thus minimizes the amount of resizing needed to perform the stack. Finally, we reject clusters whose distance is less than the sum of their respective $R_{500}$ expressed in stack pixels, i.e. $R_{500,1}+R_{500,2}<50\text{px}$. These are likely systems lying almost along with the line of sight, such that their halos ave overlapping, in which case the bridge signal would be only sampled with few ACT beams. We do not further select clusters based on their angle with respect to the line of sight because it is not trivial to disentangle their relative motion due to the Hubble flow from their peculiar velocity from redshift measurements only. The final number of maps to be stacked is $86$. 
The stack of the rotated maps is performed by coadding them in Fourier space. Each map is deconvolved with a Gaussian beam with the nominal Compton-$y$ map FWHM of \SI{1.6}{'} scaled by the same rescaling factor applied to the map. Maps are weighted by noise level adjusted by the scale factor and the average S/N of the two clusters. The noise level is estimated by taking the standard deviation of the four corners of each map, after a $3\sigma$-clip to mask point sources and bright large scale structures, while the S/N is estimated by the matched filter used to build the cluster catalogue.
The stacked map is shown in the left panel of figure~\ref{fig:realStacks}. 
In order to better estimate the amplitude of the stacked bridge and to disentangle its emission from cluster outskirts, we fitted the 2gNFW$_{\mathrm{sph}}$+cyl-$\beta$ model and compared it to 2gNFW$_{\mathrm{sph}}$ (see section~\ref{sec:phys_models}).  Additionally, we ran a null test with non-physical cluster pairs to rule out a contamination introduced by the stacking pipeline or by uncorrelated sources (section~\ref{sec:null_test}).  

\subsection{Null test and photometric redshift clusters}
\label{sec:null_test}
\begin{figure}
    \centering
    \includegraphics[width=\columnwidth]{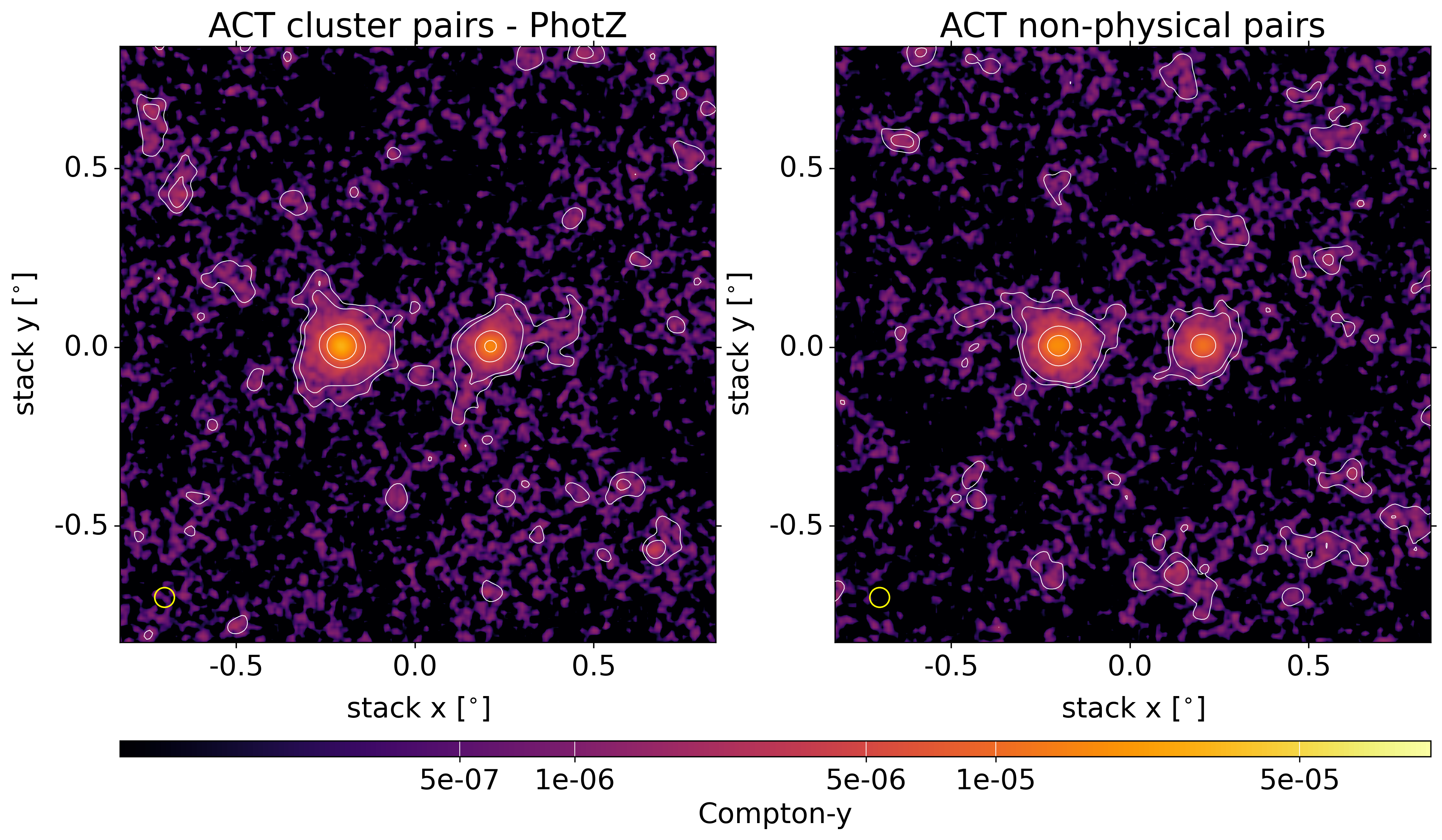}
    \caption{Left: oriented stack of 90 clusters with photometric redshift only; Right: oriented stack of 90 non-physical pairs for the null test. Contours and ACT beam as in figure~\ref{fig:realStacks}.}
    \label{fig:nullStacks}
\end{figure}

We performed a null test by repeating the same stacking procedure on a catalogue of non-physical galaxy cluster pairs, i.e., clusters that are close together on the plane of the sky ($d_{\mathrm{sky}} < 10$\,Mpc), but with a large separation along the line of sight. We selected $90$ cluster pairs from the ACT-DR6v0.1 catalogue, separated less than \SI{90}{'} on the plane of the sky and 50--200 comoving \SI{}{Mpc} along the line of sight. We verified that the mass distribution of these selected clusters is consistent with the one of the stacked pairs. The projected sky distance was chosen to reproduce the average separation of the catalogue, and match the average rescaling factor in the stack. The line of sight threshold is similar to the one used in De Graaff et al.\ \cite{de_Graaff_2019}, except with a stricter lower bound. The absence of signal between the stacked clusters shows that the excess signal between the clusters is not caused by the anisotropic stacking algorithm or random uncorrelated sources in the fields of the cluster pairs (figure~\ref{fig:nullStacks}, right).
To further justify our choice to only consider cluster pairs with spectroscopic redshift, we create the stack of only the systems with photometric redshift (figure~\ref{fig:nullStacks}, left). The photometric cluster sample is disjoint from the spectroscopic sample, and follow the same mass distribution. Photometric cluster's $5,50,95th$ quantiles are $[2.0, 2.9, 6.5]\times 10^{14}~\msun$. The intercluster signal is below the noise level of the map, suggesting that a large fraction of the cluster pairs selected with the photometric redshift are not interacting. They would thus only add noise if we included them with spectroscopically-selected pairs in a stack.

\subsection{Bootstrap test}
\label{sec:bootstrap}
We also ran a bootstrap test to determine if the observed bridge signal is dominated by a few, bright cluster systems or if it is a fainter signal common to most cluster pairs. We run the stack 100 times, each time assigning random weights to each map before the stack, extracted from a Poisson distribution with $\lambda=1$. The standard deviation measured within a $20\times\SI{20}{px}$ in the bridge region is $\sigma_{y}=3.9\times10^{-7}$, which is $\sim30\%$ of the Compton-$y$ signal integrated over the same box. We also measure the standard deviation in 66 additional non-overlapping boxes far from the stacked clusters, to verify if the observed scatter in the inter-cluster region can be completely explained by noise fluctuations, or if an additional source of scatter is required. The average standard deviation of the map measured over $20\times\SI{20}{px}$ boxes is $\sigma_{A_{\mathrm{fil}}}^{\mathrm{ACT}} = (3.31\pm0.14)\times 10^{-7}$. The scatter measured between the clusters is a $4.2\sigma$ outlier, thus the observed fluctuations are unlikely to be only caused by noise.

\begin{figure}
    \centering
    \includegraphics[width=\columnwidth]{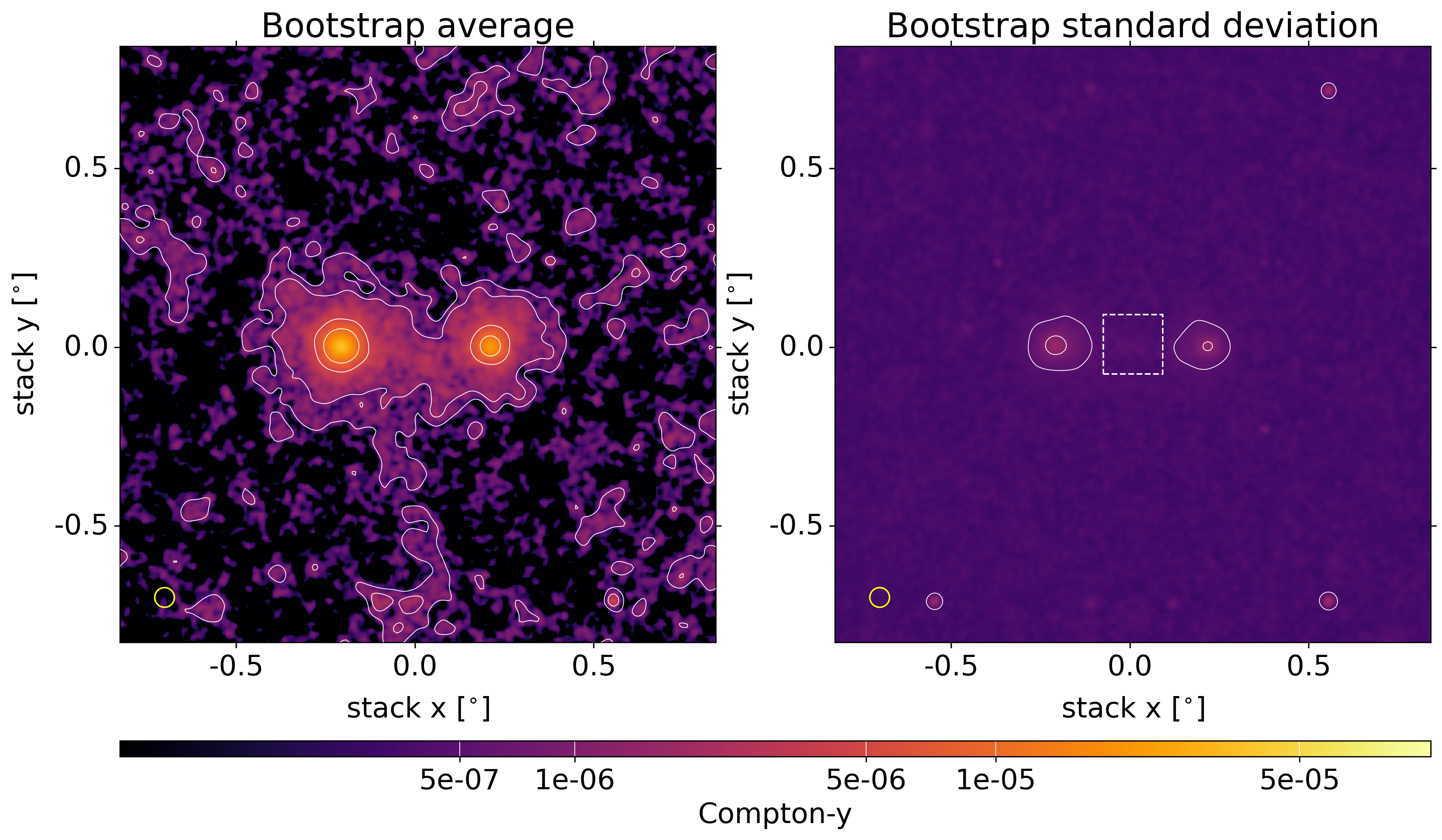}
    \caption{Left: average of 100 bootstrapped maps; Right: standard deviation of 100 bootstraped maps. Contours and ACT beam as in figure~\ref{fig:realStacks}. The white dashed box shows the $20\times\SI{20}{px}$ used to estimate the standard deviation to be compared with simulations in section~\ref{sec:sims}. }
    \label{fig:boostrap}
\end{figure}

\subsection{Model and MCMC fit}
\label{sec:stack_mcmc}
In order to estimate the Compton-$y$ signal in the intercluster region we fit the stacked map with the same approach as the single pairs (see section~\ref{sec:single_pairs}). Clusters are modeled with a gNFW profile and the bridge is modeled with a cylindrical $\beta$-model (see section~\ref{sec:phys_models}), with the difference that the amplitude is not given by $P_0$ but by the Compton-$y$ amplitude $A_{\mathrm{fil}}$. We test different models, and use the likelihood ratio and the Akaike Information Criterion to find the best fit for our data.

The main difference with respect to the single pairs is that we leave the $l_{fil}$ free, instead of the single pairs, in which we fixed the size of the bridge and only left its amplitude $A_{fil}$ free.

The models used for the fit and the relevant statistics are listed in table~\ref{tab:stat_stacks0}, while the best fit parameters are shown in table \ref{tab:fit_results_all}. We impose the condition that the filament be oriented along the axis between the clusters because of the symmetry of the stack. For the oriented stack, we initially left the $y$ coordinate free, since at least visually the bridge appeared to be not perfectly aligned with the axis between the clusters. Although the stacking symmetry should constrain the filament to have the same $y$ coordinate as the clusters, noise fluctuations can shift the peak of the signal by $1-2\SI{}{'}$. Due to the low signal-to-noise ratio of the stacked map, the addition of a free parameter to capture this shift in the MCMC lowered the statistical preference for the bridge model. We thus fixed the $y$ coordinate of the bridge to be on the intercluster axis.

Note that in general the cluster pairs are not aligned with the plane of the sky, and could have different angles with respect to the line of sight. When we model the bridge, we assume that its radial profile is conserved.

We observe an excess Compton-$y$ signal of $A_{\mathrm{fil}}=7.24^{+2.48}_{-2.29}\times 10^{-7}$, with a significance of $3.3\sigma$. Although the fit was converged, we notice a peculiar, multi-moded posterior on the filament length (figure~\ref{fig:cornerplot_stack}), which seems to converge to values larger than the distance between the clusters. This is indeed confirmed by qualitatively observing the right panel of figure~\ref{fig:realStacks}. First, the bridge does not appear particularly offset with respect to the inter-cluster axis. This supports the choice of fixing the $y$ coordinate of the bridge. Second, the bridge seems to overshoot the centers of the stacked clusters (marked with white crosses), a behaviour that was already tentatively observed in De Graaff et al.~\cite{de_Graaff_2019}. The presence of brighter clumps, with an amplitude compatible with noise fluctuations, might have caused the multi-modal posterior. Since the amplitude of the bridge does not show any correlation with its length, we proceed with the analysis without imposing any additional priors. 

\begin{table}
    \centering
    \begin{tabular}{ccccccc}%
    \hline\hline
     Model & \# free parameters & \multicolumn{3}{c}{Likelihood ratio} & $\Delta$AIC \\
    \cmidrule(lr){3-5}
    &  & $W$ & $p$-value & $\sigma$  & \\
    \hline\hline

    2gNFW$_{\mathrm{sph}}$             & 10 & --    & --                  & --  & 10.29 \\ \\
    2gNFW$_{\mathrm{sph}}$+cyl-$\beta$ & 13 & 16.29 & $9.9\times10^{-4}$ & 3.29 & 0 \\ \\
    \hline
    
    \end{tabular}
    \caption{Statistical tests for the fits of the stack. The models adopted in the fits are indicated in the first column. The number of free parameters for each model are in the second column. $W$ is the likelihood ratio (3th column), $p$-value is the probability of the null hypothesis, i.e. the simpler model with less free parameters is true (4th column), and $\sigma$ is the significance of the rejection of the null hypothesis (5th column). $\Delta_{\text{AIC},i}$ is the difference of AIC statistics with respect to the preferred model (i.e. $\Delta_{\text{AIC},i}=0$) and the other one (6th column). }
    \label{tab:stat_stacks0}

\end{table}

\begin{table}
    \centering
    \begin{adjustbox}{width=1\textwidth}
    \begin{tabular}{cccccccccc}%
    \multicolumn{10}{c}{ACT fit results}\\\\
    \hline\hline
    \multicolumn{10}{c}{Clusters}\\\\
    $x_{c1}$(px) & $y_{c1}$(px) & $x_{c2}$(px) & $y_{c2}$(px) & $\beta_{c1}$ & $\beta_{c2}$ & $A_{c1}$  [$y\times10^{-5}$] & $A_{c2}$  [$y\times10^{-5}$]& $r_{s,c1}$ [\SI{}{'}] & $r_{s,c2}$ [\SI{}{'}] \\\\
    \hline\\
    $38.23_{-0.04}^{+0.04}$ & $63.32_{-0.04}^{+0.04}$ & $88.25_{-0.04}^{+0.04}$ & $63.24_{-0.04}^{+0.04}$ & $3.32_{-0.08}^{+0.10}$ & $3.13_{-0.13}^{+0.15}$ & $6.08_{-0.15}^{+0.15}$ & $3.66_{-0.65}^{+0.18}$ & $1.53_{-0.06}^{+0.08}$ & $1.29_{-0.07}^{+0.10}$ \\\\
    \hline\hline
    \multicolumn{10}{c}{Bridge}\\\\
    & & & & $l_{\mathrm{fil}}$ [\SI{}{'}] & $r_c$ [\SI{}{'}] & $A_{\mathrm{fil}}$  [$y\times10^{-7}$] & & & \\\\
    \hline\\
    & & & & $45.61_{-9.61}^{+3.17}$ & $8.99_{-0.98}^{+3.64}$ & $7.24_{-2.29}^{+2.48}$ & & & \\\\
    \hline\hline
\\


    \multicolumn{10}{c}{\thethree fit result}\\\\
    \hline\hline
    \multicolumn{10}{c}{Clusters}\\\\
    $x_{c1}$(px) & $y_{c1}$(px) & $x_{c2}$(px) & $y_{c2}$(px) & $\beta_{c1}$ & $\beta_{c2}$ & $A_{c1}$  [$y\times10^{-5}$] & $A_{c2}$  [$y\times10^{-5}$]& $r_{s,c1}$ [\SI{}{'}] & $r_{s,c2}$ [\SI{}{'}] \\\\
    \hline\\
    $38.23_{-0.04}^{+0.04}$ & $63.32_{-0.04}^{+0.04}$ & $88.25_{-0.04}^{+0.04}$ & $63.24_{-0.04}^{+0.04}$ & $3.96_{-0.0.2}^{+0.02}$ & $4.26_{-0.07}^{+0.07}$ & $8.35_{-0.04}^{+0.04}$ & $2.70_{-0.05}^{+0.05}$ & $2.17_{-0.02}^{+0.02}$ & $2.23_{-0.0.07}^{+0.06}$ \\\\
    \hline\hline
    \multicolumn{10}{c}{Bridge}\\\\
    & & & & $l_{\mathrm{fil}}$ [\SI{}{'}] & $r_c$ [\SI{}{'}] & $A_{\mathrm{fil}}$  [$y\times10^{-7}$] & & & \\\\
    \hline\\
    & & & & $20.13_{-0.79}^{+0.62}$ & $4.54_{-0.48}^{+0.52}$ & $3.32_{-0.34}^{+0.35}$ & & & \\\\
    \hline\hline
\\


    \multicolumn{10}{c}{MAGNETICUM fit results}\\\\
    \hline\hline
    \multicolumn{10}{c}{Clusters}\\\\
    $x_{c1}$(px) & $y_{c1}$(px) & $x_{c2}$(px) & $y_{c2}$(px) & $\beta_{c1}$ & $\beta_{c2}$ & $A_{c1}$  [$y\times10^{-5}$] & $A_{c2}$  [$y\times10^{-5}$]& $r_{s,c1}$ [\SI{}{'}] & $r_{s,c2}$ [\SI{}{'}] \\\\
    \hline\\
    $39.8_{-0.01}^{+0.01}$ & $63.80_{-0.01}^{+0.01}$ & $88.77_{-0.03}^{+0.03}$ & $63.80_{-0.03}^{+0.03}$ & $4.43_{-0.03}^{+0.03}$ & $3.35_{-0.04}^{+0.04}$ & $6.19_{-0.03}^{+0.03}$ & $2.20_{-0.04}^{+0.04}$ & $4.01_{-0.05}^{+0.05}$ & $1.99_{-0.05}^{+0.07}$ \\\\
    \hline\hline
    \multicolumn{10}{c}{Bridge}\\\\
    & & & & $l_{\mathrm{fil}}$ [\SI{}{'}] & $r_c$ [\SI{}{'}] & $A_{\mathrm{fil}}$  [$y\times10^{-7}$] & & & \\\\
    \hline\\
    & & & & $36.48_{-0.35}^{+0.36}$ & $3.44_{-0.28}^{+0.26}$ & $8.17_{-0.41}^{+0.39}$ & & & \\\\
    \hline\hline

    \end{tabular}
    
    \end{adjustbox}
    \caption{Best-fitting values for the free parameters adopted in the 2gNFW$_{\mathrm{sph}}$+cyl-$\beta$ model for the real and simulated stacks. Subscript 'C1' ('C2') refers to the cluster in the left (right) side of the stacked $y$-map (see figure~\ref{fig:realStacks}). The values correspond to the medians of the posterior distributions, with $1\sigma$ errors referring to 16th and 84th percentiles. The cluster coordinates $x$ and $y$, and the radii $r_s$ are in arcminutes, while the amplitudes $A$ are in units of Compton-$y$ parameter. }
\label{tab:fit_results_all}

\end{table}

\section{Comparison with simulations}
\label{sec:sims}
We compare our stacked results with hydrodynamical simulations from \thethree\footnote{https://www.nottingham.ac.uk/astronomy/The300/index.php} and Magneticum Pathfinder.\footnote{http://www.magneticum.org/} The main difference between the two simulation sets is that \thethree is a set of zoomed-in simulations of massive clusters, while Magneticum Pathfinder is a full cosmological hydrodynamic simulation. In both cases, we create mocks of the ACT observations by selecting cluster pairs with the similar properties as the catalogue we are using, and then stacking them with the same pipeline used for ACT data. To neglect projection effects, we used the same line of sight projection length \texttt{Z\_KPC}$=\texttt{0.15e5}=\SI{15}{Mpc}$ for both simulation suites.

\subsection{THE300 project}\label{sec:thethreehundred}

\begin{figure}
    \centering
    \includegraphics[width=\textwidth]{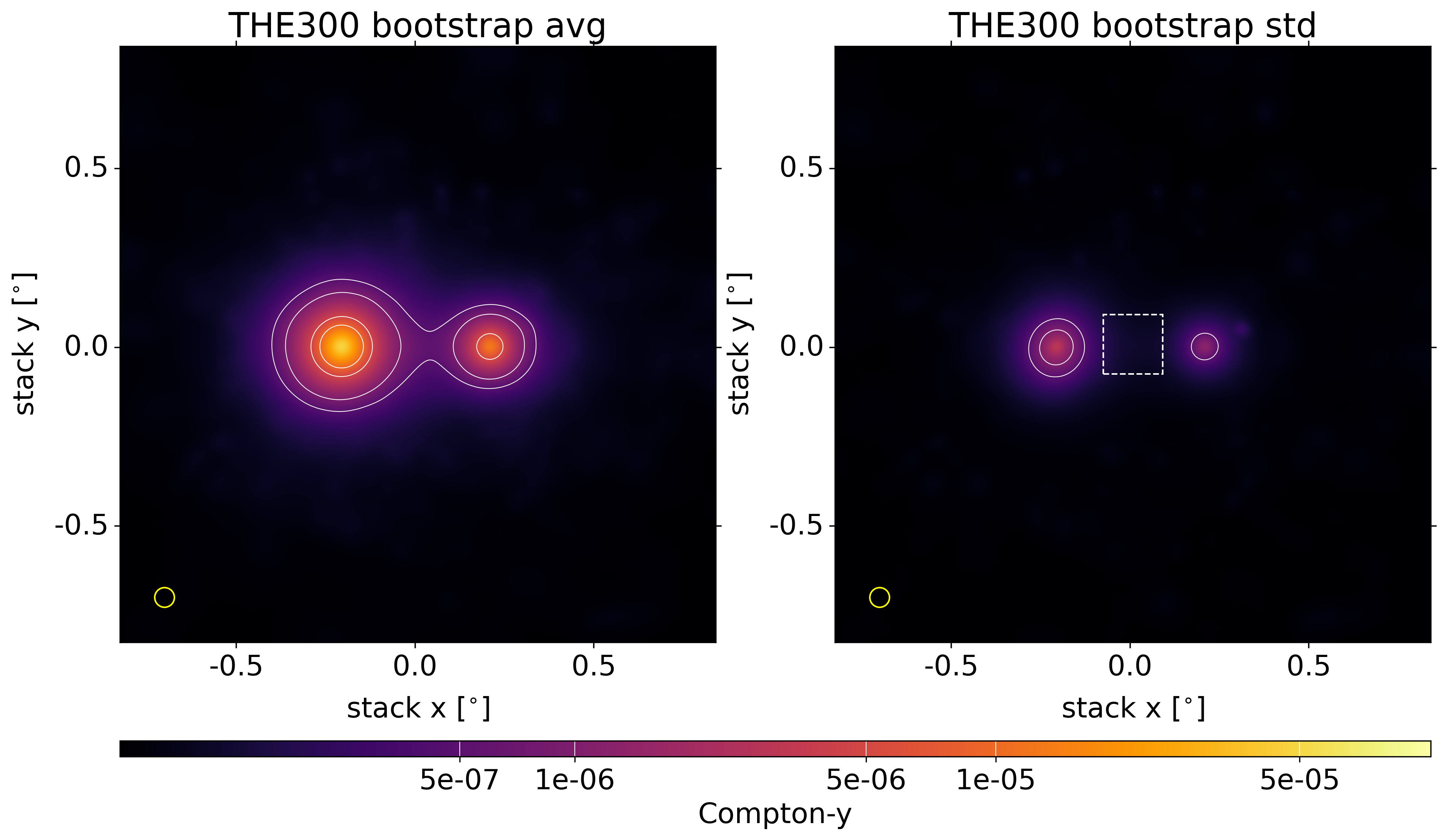}
    \caption{Stack of simulated maps from \thethree, as described in section~\ref{sec:thethreehundred}. The left panel shows the average of 100 bootstrap realizations, while the right panel shows the standard deviation. The white dashed box is the $20\times20\SI{}{px}$ area used for estimating the uncertainty of the best fit filament amplitude. Contours are the same as figure~\ref{fig:realStacks}. 
    }
    \label{fig:the300stack}
\end{figure}

\thethree simulation is composed of 324 large galaxy clusters \cite{Cui2018}, re-simulated at high resolution using the GADGET-X smoothed-particle hydrodynamics code, starting from the DM-only MDPL2 simulation, which simulates a periodic cube with a volume of 1~Gpc$^3$ with a resolution of $1.5\times10^9~\msun/h$ assuming Planck15 cosmology \cite{planck2016}. For each cluster, re--simulated into $(\SI{30}{Mpc/h})^3$ boxes with a resolution of $2.36\times10^{-8}/h \SI{}{\msun}$, we extracted three maps by projecting 18 redshift snapshots at $0.02<z<1.96$ onto the $xy$, $xz$ and $yz$ planes. In each projected map, cluster pairs were selected with the same criteria of projected distance and peculiar velocity used to compile the ACT Multiple Cluster Catalogue. Considering all snapshots and selected pairs, we obtain a total of 14,311 maps. We rejected ``crowded" maps, i.e., maps with more than two halos with a peak above $y=10^{-5}$, in order to avoid stacking systems with more than two components, similarly to what is done in the ACT stack. Additionally, we reject maps in which the clusters were overlapping, i.e., the distance between the two clusters is less than $R_{500,1}+R_{500,2}$. The final sample contains 1,049 maps. In order to match the real sample, we resampled the catalogue. First, we divided the ACT catalogue into four redshift bins and four mass bins, for a total of 16 bins. We then extracted a proportional number of maps from the simulated sample to reproduce the observed distribution. We built 100 stacks, each one of them composed of 100 randomly chosen maps, which is a number of maps close to
the number in the stack presented in section~\ref{sec:stack}. We then computed the average and the standard deviation maps, shown in figure~\ref{fig:the300stack}.

\subsection{Magneticum Pathfinder} \label{sec:magneticum}
\begin{figure}
    \centering
    \includegraphics[width=\textwidth]{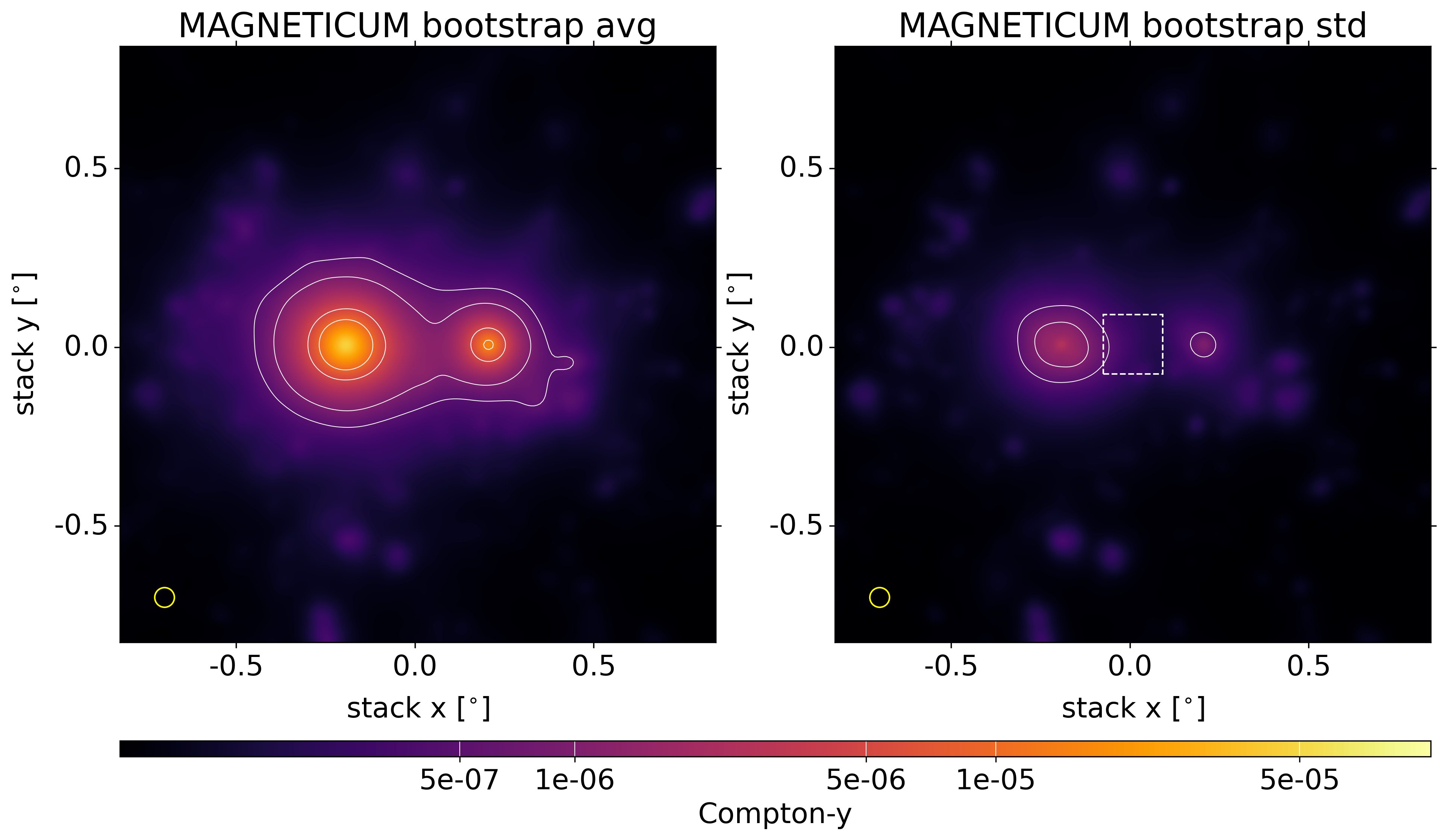}
    \caption{Stack of simulated maps from Magneticum, as described in section~\ref{sec:magneticum}. Panels and contours are the same as the previous figure.
    }
    \label{fig:stack_magneticum}
\end{figure}

The Magneticum Pathfinder suite is composed of hydrodynamical simulations of different cosmological volumes, each of them sampled with a large number of particles to provide better spatial resolution. Simulations were performed with the Tree/SPH code Gadget-3, a refined version of Gadget-2 code \cite{Springel_2005}. In this work we used the \textit{hr} resolution run of \textit{box2b}, with a volume of ($640$ comoving Mpc/h)$^3$, containing $2 \cdot 2880^3$ particles. The box was simulated using a $\Lambda$CDM cosmological model with WMAP7 parameters \cite{Komatsu_2011} and includes a number of physical and astrophysical processes, including star formation and chemical evolution, black holes and AGN feedback as well as large-scale magnetic fields.  A sample of $\approx 2000$ clusters with mass similar to the one observed by ACT was extracted from seven snapshots at redshift $0.25 < z < 0.9$. Then, a list of cluster pairs was compiled using the cluster centers provided by the code outputs, identified using the SubHalo halo finder algorithm \citep{Dolag2009}. The full cluster catalogue was then parsed to identify pairs matching our criterion. Simulated Compton-$y$ maps were subsequently computed using \textsc{Smac}, a map-making utility for idealized observations \cite{Dolag_2005}. Maps were centered on the (projected) cluster bridge, and integrated along the line of sight in the same way as for THE300 sample. Cluster pairs were initially identified using the same criteria of peculiar velocity difference and projected distance as Hilton et al.\ \cite{Hilton_2021}. Then, crowded maps and maps with overlapping clusters were rejected with the same methodology as used for \thethree. The remaining maps were divided into mass and redshift bins, and then resampled and bootstrapped in the same was as for the \thethree sample (section~\ref{sec:thethreehundred}). The result is shown in figure~\ref{fig:stack_magneticum}.


\subsection{Modeling}

We fitted the simulated maps with the same model as the ACT stack (section~\ref{sec:stack_mcmc}). Unlike the ACT maps, the simulated maps are noiseless. However, small halos around the main clusters create a Poisson-like noise in the simulated stack that is not accounted for in our MCMC likelihood, which assumes the noise is Gaussian. We therefore needed to inject noise into the map, such that it matches the real observation. The ACT noise is complex, because it is a combination of white noise and large scale noise from atmospheric emission. Moreover, there's an additional contribution from the component separation algorithm used to build the Compton-$y$ map. A possible approach could be stacking empty parts of the Compton-$y$ map and add noise that is, indeed, real data. However, for better control, we proceed with a simple Gaussian noise, with the aim of building a dataset from simulated data only. We thus injected random, Gaussian noise model smoothed to the \SI{1.6}{'} ACT+\Planck $y$-map resolution and matched to the noise level of the $y$-map ($\sigma_y\approx7\times 10^{-5}$). 
 We estimate the uncertainty of the bridge amplitudes by computing the standard deviation of the noiseless bootstrap maps (see Sections \ref{sec:thethreehundred} and \ref{sec:magneticum}) in a $20\times20\text{px}$ box in the center of each stack. This method gives us uncertainties of $\sigma_{A_{\mathrm{fil}}}^\text{MAG}=2.92\times 10^{-7}$ for Magneticum and $\sigma_{A_{\mathrm{fil}}}^\text{THE300}=1.36\times 10^{-7}$ for \thethree. We observe that both simulations are consistent with the observations: $3.32\times10^{-7}$  and $8.17\times10^{-7}$ which, when compared to our measurement, are 1.7 sigma low and 0.4 sigma high.  We will further comment on the factor of ~2.5 difference in the filament tSZ amplitude between the simulations in the discussion section.  The full results of the fits can be found in table~\ref{tab:fit_results_all}. The results more relevant for the discussion are summarized in table~\ref{tab:sims_fit_results}. For the corner plots of the mcmc, see appendix~\ref{apx:sims_fit_results}.


\begin{table}[]

    \centering

    \begin{tabular}{cccccc}
    \hline
    \hline\\
    Simulation      &   $A_{\mathrm{fil}}\times10^{-7}$   &  $\sigma_{A_{\mathrm{fil}}}^{\mathrm{mcmc}}\times10^{-7}$ & $\sigma_{A_{\mathrm{fil}}}^ {\mathrm{boot}}\times10^{-7}$& $r_c$ [\SI{}{'}] & $l_{\mathrm{fil}}$ [\SI{}{'}] \\
    \\
    \hline
    \hline\\
    \thethree          &   $3.32$  &  $0.35$  &  $1.36$   & $4.54\pm0.70$ &  $20.1\pm0.7$    \\
    Magneticum      &   $8.17$  &  $0.4$  &  $2.92$   & $3.44\pm0.25$ &   $36.5\pm3.6$   \\
    \\
    \hline
    \end{tabular}
    \caption{Best fit results for \thethree and MAGNETICUM simulations. We only show the filament parameters that are relevant for the discussions; see appendix~\ref{apx:sims_fit_results} for the full results.}
    \label{tab:sims_fit_results}
\end{table}

\section{Discussion}
\label{sec:discussion}
In section~\ref{sec:stack}, we reported a measurement of the Compton-$y$ signal measured between massive ($\left<M_{500}\right>\approx 3.5\times10^{14}\,\msun$) clusters detected in the preliminary ACT-DR6v0.1 cluster catalogue. We measure an excess of $y=(7.24^{+2.48}_{-2.29})\times 10^{-7}$ with a significance of $3.3\sigma$. 

If we calculate the $R_{500}$ of the stacked clusters as $r_s \times c_{500}$ assuming $c_{500}=1.81$ \citep{PlanckClusterProfiles2013}, and then convert it to $R_{200}$, we observe that, for the left cluster, $R_{200}=4.2'$. This means that the center of the filament is at roughly $3\times R_{200}$. This gives us confidence that the signal we observe is located well outside the outskirts of the clusters.

We then compared the measurement with mock observations created from \thethree and Magneticum-pathfinder simulation suites (section~\ref{sec:sims}). The two estimates from simulations are compatible within $2\sigma$ with the observed Compton-$y$ value. Some degree of discrepancy is expected between the two simulations as they differ in radiative models, AGN feedback, volume and assumed cosmology. Re-simulations of the same clusters with various simulation codes and different radiative models  \cite{NIFTY_sims_comp} found discrepancies of the order of few tens of percent. In our case, however, we are observing less dense environments, which we expect to be much more sensitive to AGN feedback. This could explain the higher discrepancy we observed between the two simulations compared to literature. 
\subsection{Comparison with literature and evidence for a scaling relation.}
The inter-cluster excess we find in our stacked data and the simulations is significantly larger compared to other works studying the cosmic web by stacking on \Planck and/or ACT maps. However, past studies involved smaller halos, since they used CMASS and LRG galaxies which are typically associated with halo masses of the order of $M_{500}\approx 10^{13}\,\msun$, about an order of magnitude smaller than our sample of clusters, whose average mass is $M_{500}=3.5\times 10^{14}\,\msun$. In these previous studies \cite{de_Graaff_2019, tanimura_2019, Singari_2020}, the intercluster signal was found to be $y\approx 10^{-8}$, which is also about an order of magnitude fainter than what we find. Similar results, with a Compton-$y$ amplitude of the order of $10^{-8}$ are found when applying the DisPerSE algorithm on the same samples of galaxies to build the average radial profile of a large number of filaments \cite{Tanimura2020, MallabyKay24}.

\begin{figure}
    \centering
    \includegraphics[width=\textwidth]{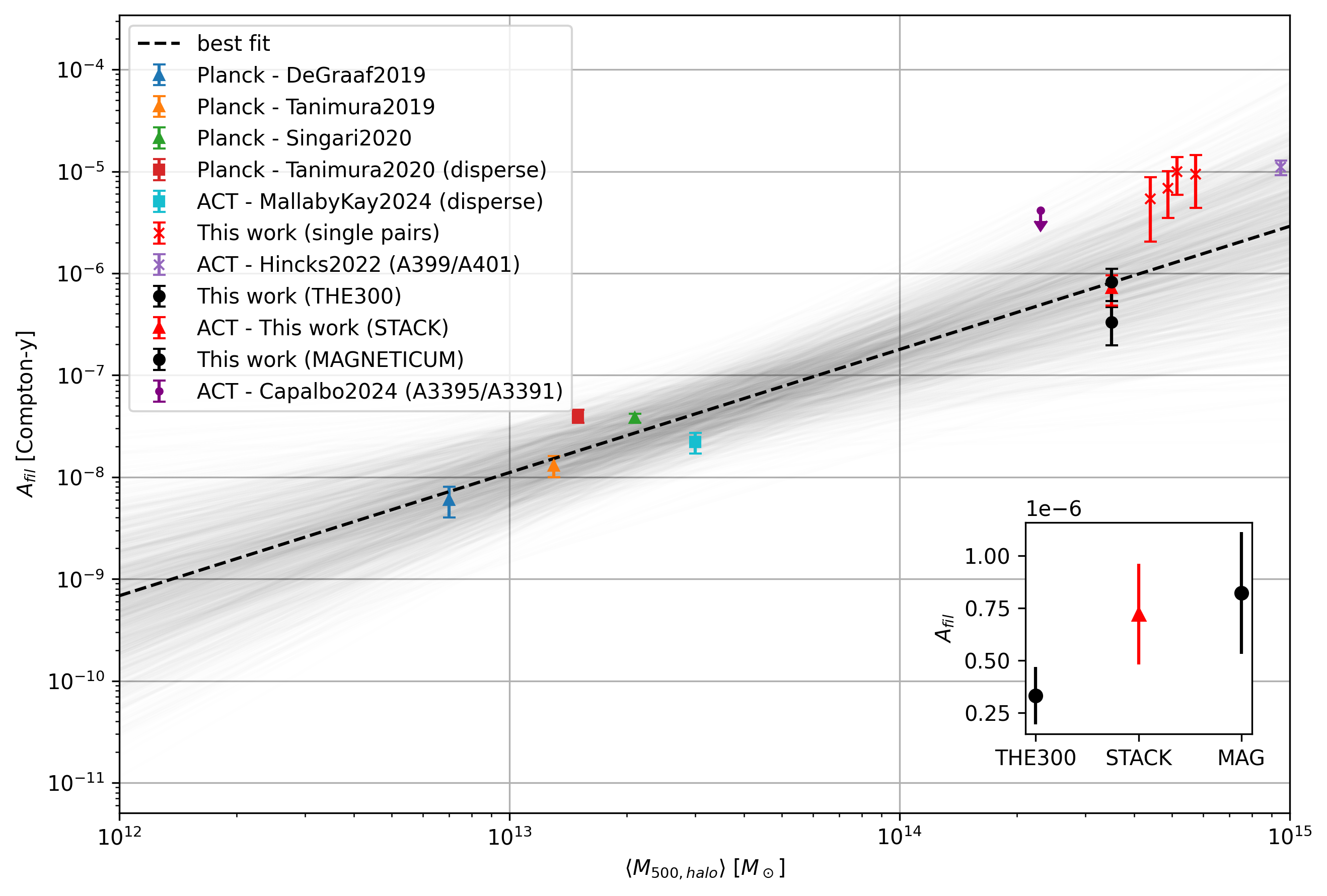}
    \caption{Measurements of SZ signal from filaments plotted in the $\braket{M_{500}}-\hat{y}$ plane, where $\hat{y}$ is the peak Compton-$y$ signal, showing the correlation between the two quantities. We plot results from pair stacking works (triangles), DisPerSE stacking works (squares) and single detections (crosses). The dashed line is the best fit performed only on the pair stacking results (triangles), and the best fit slope is $1.21^{+0.34}_{-0.35}$. Shaded grey lines shows the uncertainty. The subplot in the bottom-right corner gives a quick comparison between the stack presented in this work and the simulations described in section~\ref{sec:sims}.
    }
    \label{fig:paper_comparison}
\end{figure}

Figure~\ref{fig:paper_comparison} puts our results in the context of the previous studies mentioned before, and shows a clear scaling of the bridge amplitude with halo mass. This trend is qualitatively consistent with studies of the cosmic web in cosmological simulations, which have found evidence for a scaling relation in filaments. Galarraga-Espinosa et al.~\citep{GalarragaEspinosa2022} measured the radial profiles of filaments identified with the DisPerSE algorithm on a box from the Illustris-TNG simulation and found a positive correlation between density of the filament and its temperature, as well as between density and pressure. This correlation means that the SZ signal should be enhanced in denser filaments. Furthermore, Zhu et al.~\citep{zhu2024relation}, studying filaments in hydrodynamic simulations, linked the density, $\rho_{\mathrm{fil}}$, and width of filaments, $D_{\mathrm{fil}}$, to the mass of the halos embedded in them, although with a large scatter. In particular, $\log{D_{\mathrm{fil}}}\propto \log{n_{\mathrm{fil}}}$, while the density scales approximately as $D_{\mathrm{fil}}^{2}$. The combination of the former relation with the latter shows that filaments with massive halos will be denser, hotter and wider. The combination of our results with the studies stacking on lower density WHIM (figure~\ref{fig:paper_comparison})  is evidence that the scaling relation reported in simulations exists in reality: filaments embedding massive clusters are denser, hotter and wider, thus showing an enhanced tSZ signal. We fitted the slope of the relation shown in figure~\ref{fig:paper_comparison} using only the measurements from pair stacking papers (triangle data points) and observe that the best fit slope of $1.21^{+0.34}_{-0.35}$. However, the slope we report is mainly driven by our measurement in the high mass range. Although filaments are not virialized, thus they do not show a self similar behaviour like clusters, we see that the slope is compatible with the $5/3$ slope for clusters. At this stage, with very few stacking measurements, this result should be taken as a qualitative scaling relation, that can be used for rough prediction of the expected signal between clusters of a certain mass, but it will be interesting to physically interpret this scaling relation in the future. 

Among the points plotted in figure~\ref{fig:paper_comparison} we also show the four clusters we fitted in section~\ref{sec:single_pairs}. It is interesting to note that, although the evidence for a bridge component in these systems is marginal ($\approx 2\sigma$), they seem to be consistent with other single filament detections, in the sense that they systematically appear above the best fit scaling relation. Although not significant, it might hint that the detections of single filaments are driven by observational biases, be a consequence of the fact that not all cluster pairs in our stacked sample have a bridge, biasing our estimate of $A_{fil}$ low. Another possible interpretation is that the bridges we detect are almost aligned with the line of sight. Although it is difficult to disentangle the hubble flow from the cluster's peculiar velocities, for three out of four pairs we show in \ref{sec:single_pairs}, their $\Delta z$ is rather significant, hinting that we might be observing them at an acute angle, similarly as the A399--A401 system (as found in Hincks et al. (2022) \cite{Hincks_2022}) or the A3391--3395 system (Tittley et al. 2001 \cite{Tittley2001}; note that a high resolution investigation of this system is underway with ACT data by Capalbo et al., in prep). In both scenarios, deeper data from future surveys will be crucial to better understand the observational biases that are currently affecting the detection of single filaments. 
\subsection{Estimate of the overdensity}
Another interesting quantity to be compared with other articles is an estimate of the overdensity of the filament we observe. The tSZ data, however, only allow us to constrain the pressure, or the product between density and temperature (eq.~\ref{eqn:tSZ}). In order to break the degeneracy, we would need additional X-ray or lensing data. We defer a more detailed exploration of using lensing to a future study; here, though, we exploit two scaling relations to estimate the electron temperature of our stacked result.

\begin{figure}
    \centering
    \includegraphics[width=\textwidth]{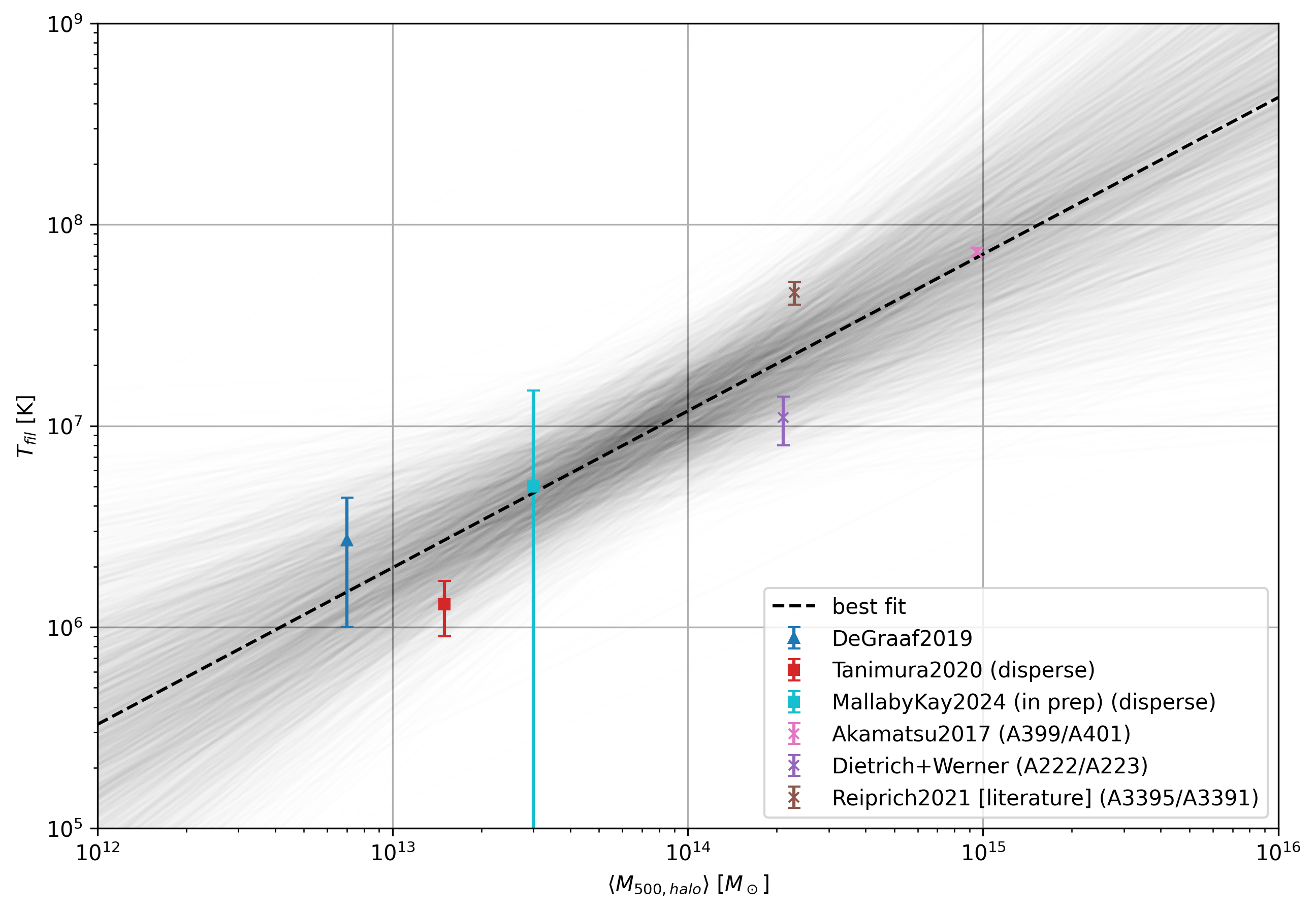}
    \caption{Measurements of electron temperature in filaments, showing a correlation between the two quantities. We plot results from pair stacking works (triangles), DisPerSE stacking works (squares) and single detections (crosses). The dashed line is the best fit performed on all the data points, and the best fit slope is $0.78^{+0.24}_{-0.24}$. Fitting on just the stacking data points gives a slope of $0.84^{+0.56}_{-0.87}$. Shaded grey lines show the uncertainty of the fit with all points included.
    }
    \label{fig:T_vs_m500}
\end{figure}

The scaling relation between linear halo mass and density reported in Zhu et al.\ \citep{zhu2024relation} shows that we can use the width of the filament, which is easier to constrain, especially with the high resolution of ACT, to break the degeneracy between temperture and density. Starting from Equation~\ref{eqn:tSZ} and using the scaling relation between $D_{\mathrm{fil}}$ and $n$ reported in Zhu et al., and assuming that $n_e\propto n$, the Compton-$y$ signal will scale as:
\begin{equation}
y\propto \int_{\mathrm{los}}n_eT_e\text{d}\ell \approx n_e T_e D_{\mathrm{fil}}\propto D_{\mathrm{fil}}^{2}T_eD_{\mathrm{fil}}\propto T_eD_{\mathrm{fil}}^{3}
\end{equation}
where $n_e$ is the electron density and $T_e$ is the electron temperature. We assume that $D_{\mathrm{fil}}\propto r_c$.
We can then combine two measurements of $D_{\mathrm{fil}}$ and $y$ in different halo mass ranges to estimate the electron temperature of our cluster sample as follows:
\begin{equation}
\frac{T_{e}^{\mathrm{hi}}}{T_e^{\mathrm{lo}}} = \left(\frac{r_c^{\mathrm{lo}}}{r_c^{\mathrm{hi}}}\right)^3\left(\frac{y^{\mathrm{hi}}}{y^{\mathrm{lo}}} \right)
\end{equation}
where ``hi'' and ``lo'' indicate the high mass and low mass halo samples, $y$ is the peak Compton-y signal of the best fit cylindrical-$\beta$ model, $T_e$ is the electron temperature and $r_c$ is the core radius of a cylindrical $\beta$ model. Mallaby-Kay et al.\ \citep{MallabyKay24} and Tanimura et al.\ \citep{Tanimura2020}, used the DisPerSE algorithm to compute the stacked Compton-$y$ and lensing $\kappa$ radial profiles of filaments identified from the 3D distribution of CMASS galaxies, using respectively ACT and \Planck data. We choose to use the result from Mallaby-Kay et al.\ since they use our same dataset: $r_c=1.8^{+0.6}_{-0.4}\,\SI{}{Mpc}$ and $T_e=(5\pm1)\times 10^6\,\SI{}{K}$. Substituting these values of $r_c$ and $T_e$ and the best-fitting $r_c$ from our sample listed in \ref{tab:fit_results_all}, we obtain an electron temperature of $(1.5\pm 0.6)\times 10^6\,\SI{}{K}$. We can also extrapolate the electron temperature to our sample with a linear extrapolation of the temperatures measured in galaxy stacking works \citep{de_Graaff_2019, MallabyKay24, Tanimura2020} (figure \ref{fig:T_vs_m500}). As stated before, we expect the temperature to scale with a slope smaller than the self-similar value of $2/3$. Current data are not sufficient to firmly constrain this index, because we do not have temperature measurements of filaments in the high mass range, except for few notable systems. If we do not include these systems, the best fit slope is $0.84^{+0.56}_{-0.87}$, which is poorly constrained and even compatible with a zero slope within $1\sigma$. Including these clusters brings the value to $0.78^{+0.24}_{-0.24}$, however, since we mix clusters at different redshifts (stacking at $z\approx 0.5$ and single pairs in the local universe), we do not use this slope. Since we cannot firmly constrain the electron temperature by fitting the mass-temperature relation, we use the theoretical slope from self similarity of clusters, $T\propto M^{2/3}$, to extrapolate the temperature measured in Mallaby-Kay et al.\ to our mass range. We stress that it is a strong assumption, since filaments are not virialized. However, one might expect that in a sub--virial regime like the one found in filaments, the slope from clusters can be taken as an upper limit. Assuming this slope, we obtain a temperature $T_e = (2.2\pm0.4)\times 10^{7}\,\SI{}{K}$. This temperature is significantly higher than the one derived from the filament thickness scaling relation. On the other hand, the small $T_e$ derived from the scaling relation in Zhu et al.\ \cite{zhu2024relation} could be underestimated because of the large scatter of that relation in the high mass range. This further emphasizes the need for better observations and simulations  of the filaments around massive halos. If we assume the former temperature estimated by combining this work and the results from Mallaby-Kay et al., we obtain a central overdensity of $\delta = 25\pm14$. The uncertainty is very large, and it is driven by the current uncertainty on $T_e$. If we use the latter temperature to estimate the overdensity, we obtain $\delta = 2.5\pm1.7$, which is about an order of magnitude lower than the previous estimate.  This large discrepancy between estimates of $T_e$, which impacts our estimate of the average overdensity $\delta$, shows the need for targeted X-ray measurements, that will provide a direct measurement of $T_e$ and lead to a more precise estimate.

\section{Conclusions}
\label{Sect:Conclusions}

In this work we used the ACT DR6v0.1 Compton-$y$ map \cite{Coulton24_ymap} to measure the elusive matter bound to dark matter bridges connecting likely interacting galaxy clusters selected from the preliminary DR6 cluster catalogue. We followed two approaches: single pairs and stacking. In the single pair approach, we used an MCMC fit, with an approach similar to Hincks et al. \citep{Hincks_2022}, applied to single systems to estimate the Compton-$y$ amplitude of the inter-cluster region. Among the 284 systems in the catalogue, we have found four systems with a bridge evidence at S/N $>2$. After the MCMC fit, two of the four clusters show evidence at a level $>2\sigma$ for a bridge component using the ACT Compton-$y$ map alone. This result shows the difficulty of the detection of single filaments between clusters with current instrument sensitivities and resolutions. Although the results are marginal, the bridge model is not ruled out in any of the cluster pairs, which are strong candidates for deep, high resolution follow-up observations with large single dish telescopes, or for multi-wavelength observations in optical and X-ray bands.

In the second part of this work, we performed a stack of multiple systems to explore their average properties. Following the approach of past studies that used \Planck Compton-$y$ maps, we scaled and aligned cluster pairs in order to place the clusters in the same position. We then stacked the maps to obtain an average map. We used the same MCMC fit to estimate the Compton-y signal of the average bridge observed in the stacked map. We measured an amplitude of $y=(7.24^{+2.48}_{-2.29})\times 10^{-7}$ by stacking $86$ cluster pairs with an average mass $\left<M_{500} \right>\approx 3.5\times 10^{14}~M_\odot $. Including a bridge component in the fit is preferred at $3.3\sigma$ compared to the null hypothesis of no bridge component. We compared this value of Compton-$y$ excess with mock observations reproduced starting from \thethree and Magneticum hydrodynamical simulations and found that the observed amplitude from ACT is consistent with the simulations. We emphasize that this is the first work based on pair stacking that directly stacks on massive SZ detected clusters, thus probing a sample of short filaments hosted between halos in a mass range that was previously unexplored, except for the notable system A399--401. When comparing this result with other results obtained with \Planck data, we observe consistency with scaling relations derived from simulations. Indeed, the stack of cluster pairs shows that a mass--SZ scaling relation exists also for filaments, where $M$ is the mass of the clusters embedded in them, and the best fit slope is $1.21^{+0.34}_{-0.35}$. We qualitatively find this slope to be compatible with the self--similar slope of clusters, $5/3$, although the virial hypotesis does not hold for filaments. At the current stage, this relation can be used for predicting the Compton-$y$ signal expected from a filament between clusters of a certain mass.
The slope of the $M_{halo}-A_{fil}$ for filaments will be interesting to study in detail, both from simulations and from future, deeper observations. Current data do not allow us to firmly constrain the electron temperature of the stacked filament, so we are not able to estimate the baryon content of filaments in our cluster sample. However, if we assume the wide range of temperatures that we can derive from scaling relations, we obtain an overdensity range $\delta=2.5-25$. This underscores the importance of a measurement of the temperature using X-ray measurements or by combining SZ and lensing. Future surveys from instruments like the Simons Observatory \cite{simonsobs:2018} or AtLAST \cite{DiMascolo2024_atlast} will increase the yield of clusters and the Compton-$y$ sensitivity, both for studying individual systems as well enabling binning in mass and redshift. For instance, the Compton-$y$ sensitivity for AtLAST is predicted to be $\sigma_y\sim 2\times 10^{-7}$. According to the scaling relation we observe, it could directly detect filaments between clusters with a mass of $\approx 10^{14}~\msun$. The increased yield of tSZ data on individual filaments, combined with the information from future X-ray surveys \cite{LEM}, will be of fundamental importance for a deeper understanding of the physics of the cosmic web and its evolution through the history of the universe.



\bibliographystyle{JHEP}
\bibliography{biblio.bib}

\acknowledgments

We thank Craig Sarazin and Bruce Partridge for useful discussions and comments while preparing this paper.

Support for ACT was through the U.S.~National Science Foundation through awards AST-0408698, AST-0965625, and AST-1440226 for the ACT project, as well as awards PHY-0355328, PHY-0855887 and PHY-1214379. Funding was also provided by Princeton University, the University of Pennsylvania, and a Canada Foundation for Innovation (CFI) award to UBC. ACT operated in the Parque Astron\'omico Atacama in northern Chile under the auspices of the Agencia Nacional de Investigaci\'on y Desarrollo (ANID). The development of multichroic detectors and lenses was supported by NASA grants NNX13AE56G and NNX14AB58G. Detector research at NIST was supported by the NIST Innovations in Measurement Science program. Computing for ACT was performed using the Princeton Research Computing resources at Princeton University, the National Energy Research Scientific Computing Center (NERSC), and the Niagara supercomputer at the SciNet HPC Consortium. SciNet is funded by the CFI under the auspices of Compute Canada, the Government of Ontario, the Ontario Research Fund–Research Excellence, and the University of Toronto. We thank the Republic of Chile for hosting ACT in the northern Atacama, and the local indigenous Licanantay communities whom we follow in observing and learning from the night sky.

We thank The Red Española de Supercomputación for granting computing time for running most of the hydrodynamical simulations of \thethree{} galaxy cluster project in the Marenostrum supercomputer at the Barcelona Supercomputing Center. 

A.~D.~Hincks acknowledges support from the Sutton Family Chair in Science, Christianity and Cultures, from the Faculty of Arts and Science, University of Toronto, and from the Natural Sciences and Engineering Research Council of Canada (NSERC) [RGPIN-2023-05014, DGECR-2023-00180].
M.~Hilton acknowledges financial support from the National Research Foundation of South Africa.
C.~Sifón acknowledges support from the Agencia Nacional de Investigaci\'on y Desarrollo (ANID) through Basal project FB210003.
W.~Cui and G.~Yepes would like to thank Ministerio de Ciencia e Innovación (Spain) for partial financial support under project grant PID2021-122603NB-C21 
L.~Di Mascolo has been supported by the French government, through the UCA\textsuperscript{J.E.D.I.} Investments in the Future project managed by the National Research Agency (ANR) with the reference number ANR-15-IDEX-01.
D.~Fabjan acknowledges financial support from the Slovenian Research Agency (research core funding no. P1-0188).
M. De Petris and A. Ferragamo acknowledge PRIN-MUR grant 20228B938N `Mass and selection biases of galaxy clusters: a multi-probe approach" funded by the European Union Next generation EU, Mission 4 Component 1  CUP C53D2300092 0006'.
J.~P.~Hughes gratefully acknowledges support from the estate of George A. and Margaret M. Downsbrough that established the Downsbrough Chair In Astrophysics at Rutgers. 
IFAE is partially funded by the CERCA program of the Generalitat de Catalunya
WC is supported by the Atracci\'{o}n de Talento Contract no. 2020-T1/TIC-19882 which was granted by the Comunidad de Madrid in Spain.

This work made use of Astropy:\footnote{http://www.astropy.org} a community-developed core Python package and an ecosystem of tools and resources for astronomy \citep{astropy:2013, astropy:2018, astropy:2022}. 

This work made use of the \texttt{emcee} Python sampler for Markov chain Monte Carlo (MCMC) \citep{Foreman_Mackey_2013}

The work presented here emerged out of the annual The300 workshop held at UAM's La Cristalera during the week July 8-12, 2024, partially funded by the 'Ayuda para la Organización de Jornadas Científicas en la UAM en el Marco del Programa Propio de Investigación y con el Apoyo del Consejo Social de la UAM'

\newpage
\appendix

\section{Contamination from the Cosmic Infrared Background}
\label{apx:cib}
\begin{figure}
    \centering
    \includegraphics[width=.7\textwidth]{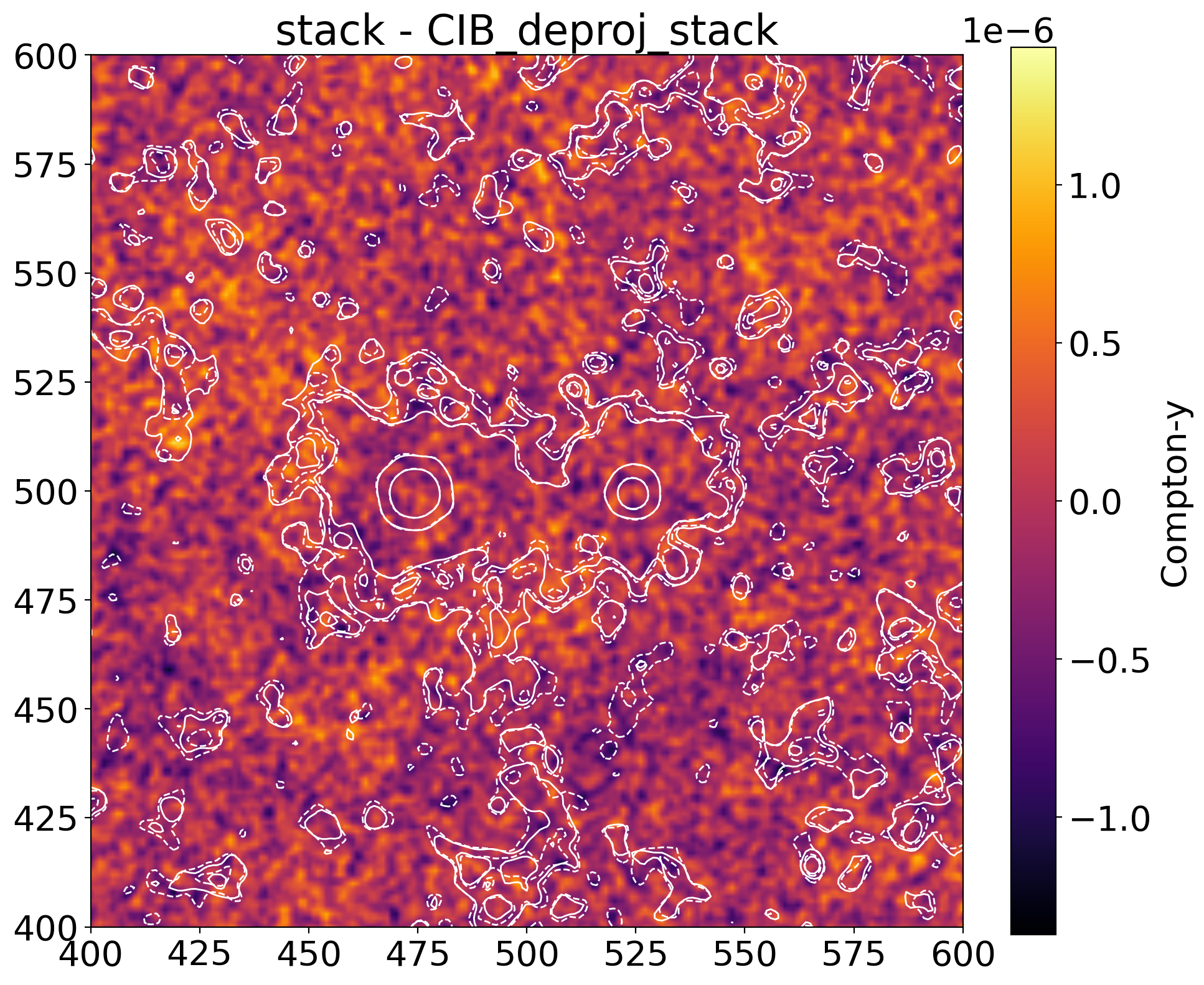}
    \caption{Residuals between the stack studied in section~\ref{sec:stack} and the same stack repeated on a CIB-deprojected version of the Compton-$y$ map. White contours show the Compton-$y$ signal in the two stacks: filled lines are for the non-CIB subtracted stack, dashed lines are for the CIB subtracted stack. 
    }
    \label{fig:cib_subtraction}

\end{figure}

The cumulative emission from unresolved dusty star forming galaxies is often referred to as the cosmic infrared background (CIB). A strong correlation has been found between the CIB emission and the tSZ signal \cite{Planck_CIB_SZ}, which is expected as both the distribution of hot gas in clusters and of galaxies depends on the underlying dark matter distribution. For this reason, we investigated the impact of the CIB on the stacking measure presented in section~\ref{sec:stack}. We ran the pipeline on a version of the Compton-$y$ map in which a dust model was included in the component separation. We then subtracted the CIB-subtracted stack from the non-deprojected stack (figure~\ref{fig:cib_subtraction}). We measured the average signal in a $50\times\SI{100}{px}$ box which encloses most of the Compton-$y$ signal from both clusters and the bridge and found an average $\Delta y_{CIB} = 3\times 10^{-8}$ which is an order of magnitude below the map noise. Therefore, we conclude that the CIB contamination is negligible, at least with our current errorbars. 

\section{Single pairs fit}
\label{apx:singlePairs}

\begin{figure*}
    \centering
    \includegraphics[scale=0.35]{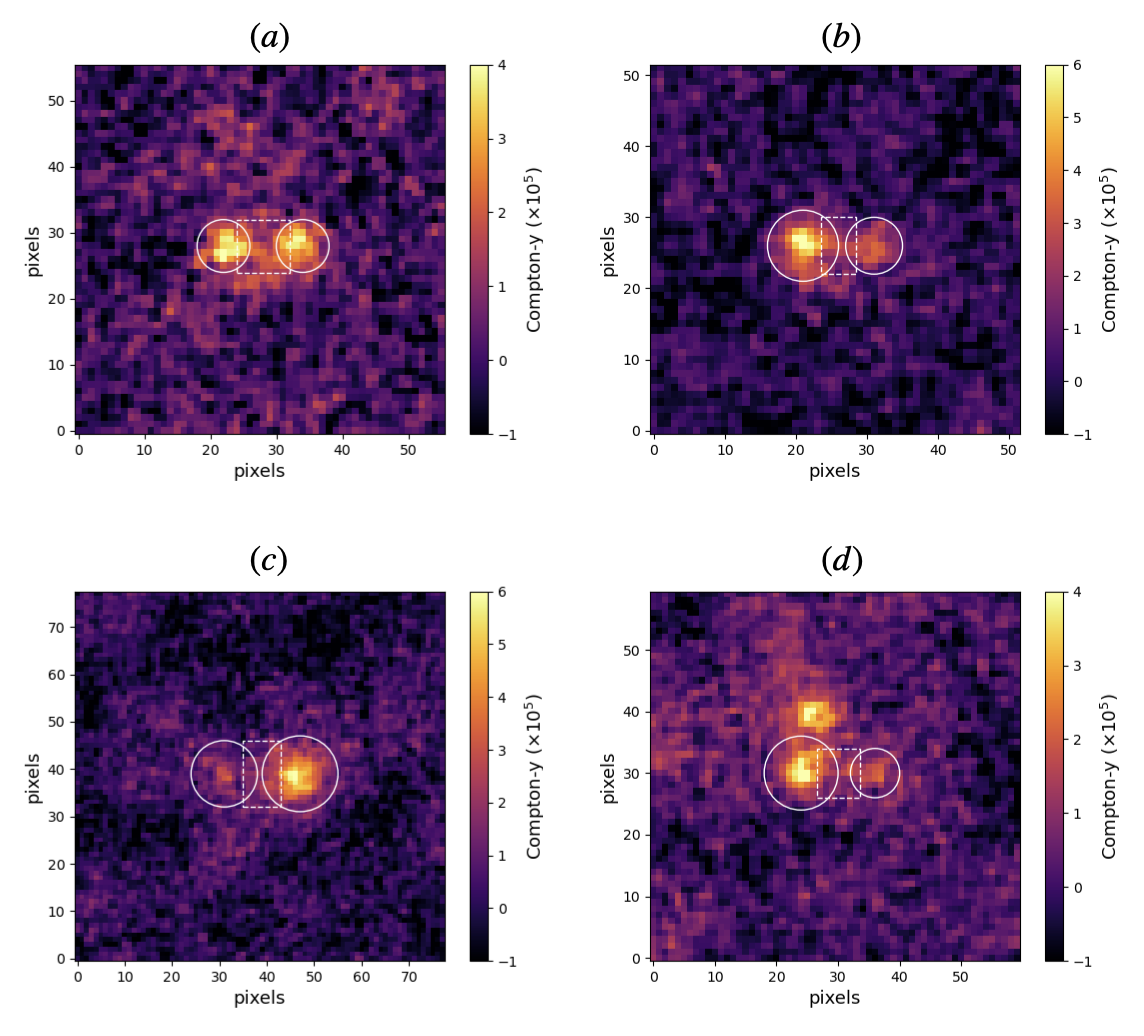}
    \caption{Compton $y$-map of selected clusters pairs (see table~\ref{tab:single_pairs}). The solid white circles are centred on the cluster coordinates and they have radii equal to $R_{500}$. The dashed boxes in white mark the regions used for the S/N estimate described in the text. The maps have a side-length of 5 times the projected separation of the clusters, with a pixel resolution of $0.5^{\prime}$. \textit{(a)}: ACT-CL~J2336.0$-$3210\,/\,ACT-CL~J2336.3$-$3206; \textit{(b)}: ACT-CL~J0328.2$-$2140\,/\,ACT-CL~J0328.5$-$2140; \textit{(c)}: ACT-CL~J0245.9$-$2029\,/\,ACT-CL~J0246.4$-$2033; \textit{(d)}: ACT-CL~J0826.0+0419\,/\,ACT-CL~J0826.4+0416.}
    \label{fig:single_pairs}
\end{figure*}

\begin{figure}
    \centering
	\includegraphics[width=.5\textwidth]{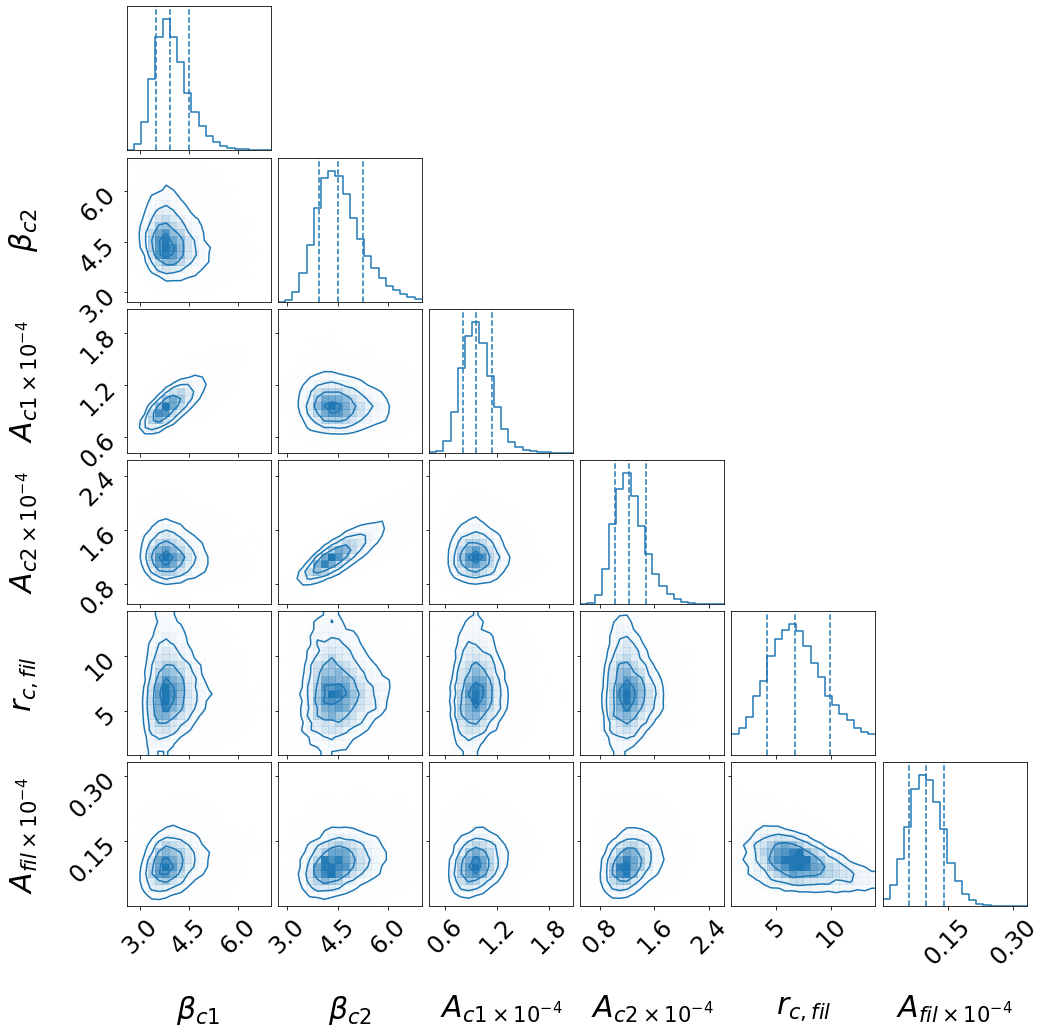}
    \caption{Posterior distributions for free parameters in the best-fitting model 2gNFW$_{\mathrm{sph}}$+cyl-$\beta$ for the cluster pair ACT-CL~J2336.0$-$3210\,/\,ACT-CL~J2336.3$-$3206. Subscript `C1' (`C2') refers to the cluster in the right (left) side of the respective $y$-map (see figure~\ref{fig:single_pairs} (a)), while `b' refer to the bridge component. Values of $A$ are in units of Compton-$y$ parameter, while $r_b$ is in arcminutes. The contours are at 16th, 50th and 84th percentiles. These are also indicated with the dashed vertical lines.}
    \label{fig:corner_plots_pair1}
\end{figure}

\begin{figure}
    \centering
	\includegraphics[width=.5\textwidth]{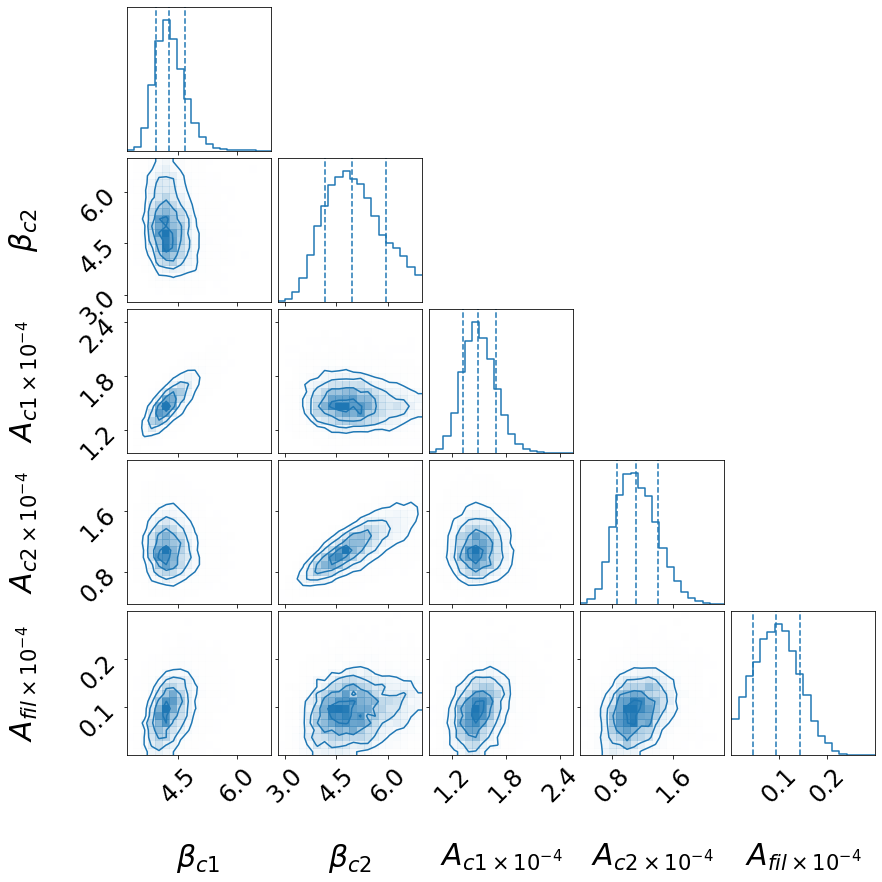}
    \caption{Posterior distributions for free parameters in the best-fitting model 2gNFW$_{sph}$ for the cluster pair ACT-CL~J0328.2$-$2140\,/\,ACT-CL~J0328.5$-$2140. Subscript ``C1'' (``C2'') refers to the cluster in the left (right) side of the respective $y$-map (see figure~\ref{fig:single_pairs} (b)). Values of $A$ are in units of Compton-$y$ parameter. The contours are at 16th, 50th and 84th percentiles. These are also indicated with the dashed vertical lines.}
    \label{fig:corner_plots_pair2}
\end{figure}

\begin{figure}
    \centering
	\includegraphics[width=.5\textwidth]{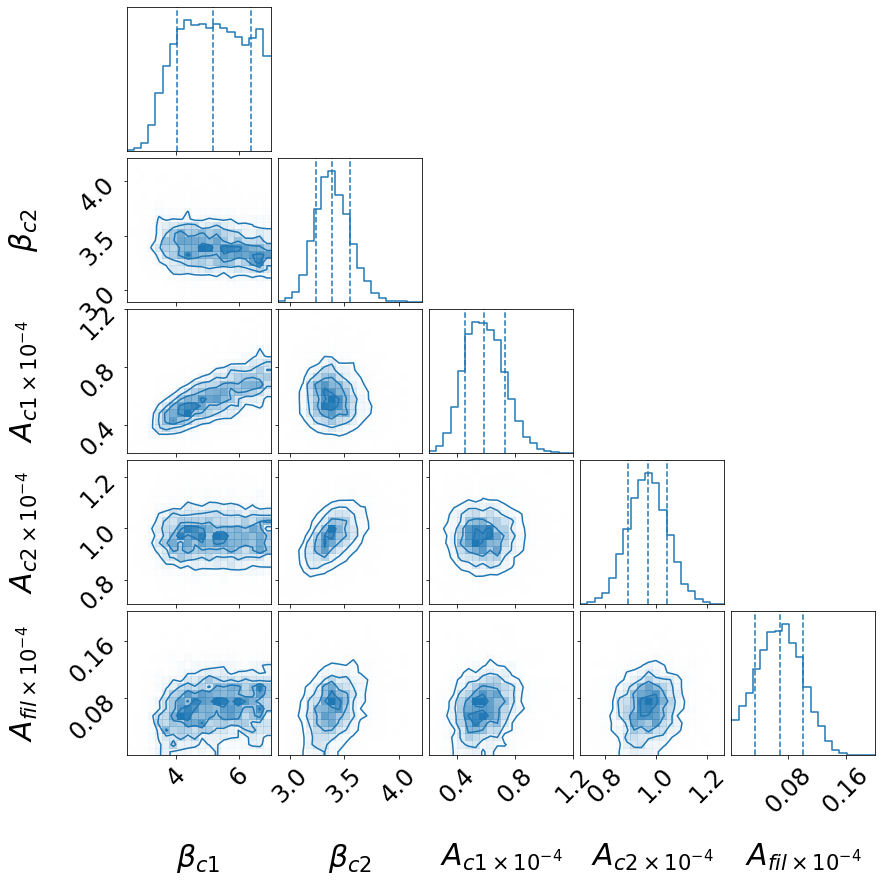}
    \caption{Posterior distributions for free parameters in the best-fitting model 2gNFW$_{sph}$ for the cluster pair ACT-CL~J0245.9$-$2029 \,/\,ACT-CL~J0246.4$-$2033. Subscript 'C1' ('C2') refers to the cluster in the left (right) side of the respective $y$-map (see figure~\ref{fig:single_pairs} (c)). Values of $A$ are in units of Compton-$y$ parameter. The contours are at 16th, 50th and 84th percentiles. These are also indicated with the dashed vertical lines.}
    \label{fig:corner_plots_pair3}
\end{figure}

\begin{figure}
    \centering
	\includegraphics[width=.5\textwidth]{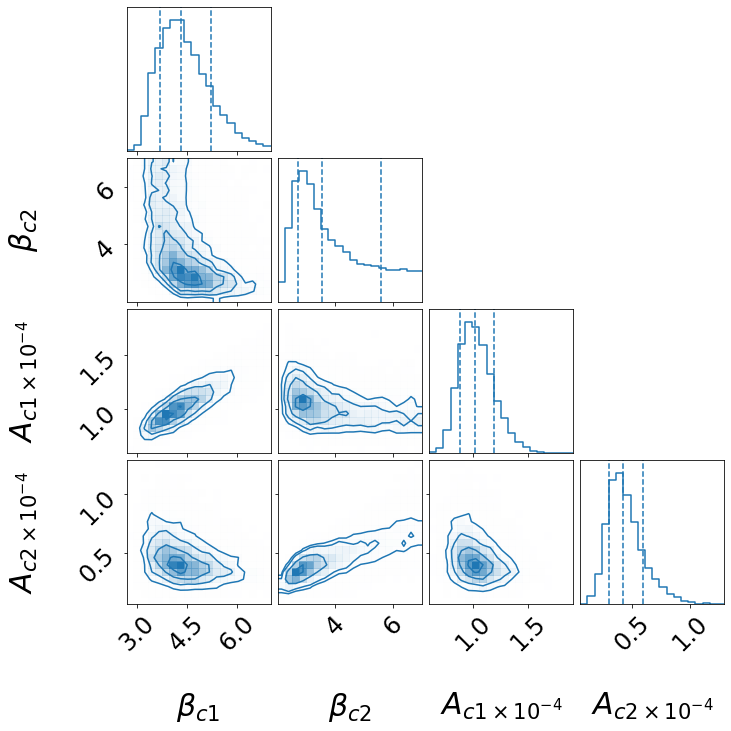}
    \caption{Posterior distributions for free parameters in the best-fitting model 2gNFW$_{sph}$ for the cluster pair ACT-CL~J0826.0+0419\,/\,ACT-CL~J0826.4+0416. Subscript 'C1' ('C2') refers to the cluster in the left (right) side of the respective $y$-map (see figure~\ref{fig:single_pairs} (d)). Values of $A$ are in units of Compton-$y$ parameter. The contours are at 16th, 50th and 84th percentiles. These are also indicated with the dashed vertical lines.}
    \label{fig:corner_plots_pair4}
\end{figure}

In this appendix we show the full results of the MCMC fit on single cluster pairs. A map of each of the selected clusters is shown in figure~\ref{fig:single_pairs}. Fit statistics and the best fitting parameters for each model mentioned in section~\ref{sec:phys_models} and \ref{sec:mcmc_fit} are summarized in table~\ref{tab:spfparams}. The corner plots showing the posterior distributions for the best fitting model of each cluster pair are presented in figures~\ref{fig:corner_plots_pair1}, \ref{fig:corner_plots_pair2}, \ref{fig:corner_plots_pair3} and \ref{fig:corner_plots_pair4}. We choose to show the best fit bridge amplitude even if it is not the best fitting model, because it is not firmly disfavoured by the statistical criteria we used, and it could be a useful indication for future follow-up proposals. Going through each fitted cluster pair, we report the following results:
\begin{itemize}
    \item ACT-CL~J2336.0$-$3210 / ACT-CL~J2336.3$-$3206 -- This pair has the highest redshift of the four selected systems, at $z_{\mathrm{pair}} = 0.62$. The redshift separation is $\Delta z = 6\times 10^{-3}$ and the sky separation is $5.5\SI{}{'}$. The statistical tests show that the best fitting model is the 2gNFW$_{\mathrm{sph}}$+cyl-$\beta$. The likelihood ratio shows a $2.3\sigma$ preference with respect to the 2gNFW$_{\mathrm{sph}}$ model. The measured excess signal has an amplitude of $y=9.9_{-3.8}^{+4.1}\times 10^{-6}$. 
    \item ACT-CL~J0328.2$-$2140 / ACT-CL~J0328.5$-$2140 -- This system has a redshift $z_{\mathrm{pair}}=0.59$ and a redshift difference of $\Delta z = 4\times 10^{-4}$. The sky separation is $5.2'$, which is the smallest of the systems we have modelled. The overlapping halos at our resolution likely contaminated the S/N estimate, which is the highest of the pairs, S/N~$\sim3.6$. Given the low S/N of the right cluster, we reduced the number of free parameters by fixing the core radius of the cylinder to the minimum $R_{500}$ of the two clusters. The statistical tests show that the preferred model is 2gNFW$_{\mathrm{sph}}$+cyl-$\beta$, and that the bridge tentative evidence is $1.4\sigma$. Nonetheless, the $\Delta$AIC value with respect to the model including only the two clusters is $<2$, indicating that the models have similar statistical support and the simpler one is not ruled out.
    \item  ACT-CL~J0245.9$-$2029 / ACT-CL~J0246.4$-$2033 -- The third pair is the lowest redshift system, with $z_{\mathrm{pair}}=0.31$, $\Delta z = 8\times 10^{-3}$ and a sky separation of $7.7'$. The signal-to-noise in the intercluster region is S/N=2.9, but as for the previous pair the secondary cluster is not firmly detected. For this reason, we fixed the core radius of the cylindrical $\beta$ model to simplify the fitting model. As in the previous case, a bridge component is neither clearly preferred nor ruled out. Infact, the $\Delta$AIC between the 2gNFW$_{\mathrm{sph}}$+cyl-$\beta$ and 2gNFW$_{\mathrm{sph}}$ is comparable with 2. The evidence for the bridge is $2.2\sigma$ for the 2gNFW$_{\mathrm{ell}}$+cyl-$\beta$ model and $1.9\sigma$ for the 2gNFW$_{\mathrm{sph}}$+cyl-$\beta$. The AIC prefers the simpler model, which is the one including circular clusters, although the bridge significance is lower.
    
    \item ACT-CL~J0826.0+0419 / ACT-CLJ0826.4+041 -- This system has a redshift $z_{\mathrm{pair}} = 0.47$ and $\Delta z = 0.01$. The two clusters are separated by $\SI{5.9}{'}$. The modeling of this system is particularly challenging because of the presence of a third foreground cluster at $z=0.2$. We proceeded to fit only the pair of interest and not to add additional complexity by modeling all three clusters. In this case, the preferred model is the 2gNFW$_{\mathrm{sph}}$. Again, the $\Delta$AIC with respect to 2gNFW$_{\mathrm{sph}}$+cyl-$\beta$ is less than $2$, thus the bridge model is not firmly disfavoured. However, the detection is not significant, at less than $1\sigma$.
    \end{itemize}
\noindent

\begin{table}

    \centering
    \begin{tabular}{cccccc}%
    \hline\hline
    \multicolumn{6}{c}{ACT-CL~J2336.0$-$3210 / ACT-CL~J2336.3$-$3206}\\ \\
    $\beta_1$ & $\beta_2$ & $A_1$  [$y\times10^{-5}$] & $A_2$  [$y\times10^{-5}$] & $r_c$ [\SI{}{'}]& $A_{\mathrm{fil}}$  [$y\times10^{-6}$] \\
    \hline\\
    $3.9_{-0.4}^{+0.6}$ & $4.5_{-0.6}^{+0.8}$ & $9.6_{-1.6}^{+1.8}$ & $12.2_{-2.0}^{+2.6}$ & $3.4_{-1.3}^{+1.6}$ & $9.9_{-3.8}^{+4.1}$ \\\\
    \hline\hline
    \multicolumn{6}{c}{ACT-CL~J0328.2$-$2140 / ACT-CL~J0328.5$-$2140}\\ \\
    $\beta_1$ & $\beta_2$ & $A_1$  [$y\times10^{-5}$] & $A_2$  [$y\times10^{-5}$] & $r_c$ [\SI{}{'}] & $A_{\mathrm{fil}}$  [$y\times10^{-6}$] \\
    \hline\\
    $4.3_{-0.3}^{+0.4}$ & $5.0_{-0.8}^{+1.0}$ & $14.9_{-1.7}^{+2.0}$ & $11.1_{-2.5}^{+2.9}$ & -- & $9.4_{-4.9}^{+4.8}$ \\\\  
    \hline\hline
    \multicolumn{6}{c}{ACT-CL~J0245.9$-$2029 / ACT-CL~J0246.4$-$2033}\\ \\
    $\beta_1$ & $\beta_2$ & $A_1$  [$y\times10^{-5}$] & $A_2$  [$y\times10^{-5}$] & $r_c$ [\SI{}{'}] & $A_{\mathrm{fil}}$  [$y\times10^{-6}$] \\
    \hline\\
    $5.2_{-1.1}^{+1.2}$ & $3.4_{-0.1}^{+0.2}$ & $5.8_{-1.3}^{+1.4}$ & $9.7_{-0.8}^{+0.8}$ & -- & $6.8_{-3.4}^{+3.2}$ \\\\
    \hline\hline
    \multicolumn{6}{c}{ACT-CL~J0826.0+0419 / ACT-CL~J0826.4+0416}\\ \\
    $\beta_1$ & $\beta_2$ & $A_1$  [$y\times10^{-5}$] & $A_2$  [$y\times10^{-5}$] & $r_c$ [\SI{}{'}] & $A_{\mathrm{fil}}$  [$y\times10^{-6}$] \\
    \hline\\
    $4.9_{-0.8}^{+0.9}$ & $4.2_{-1.3}^{+1.9}$ & $11.3_{-1.7}^{+2.1}$ & $4.9_{-1.5}^{+2.2}$ & -- & $5.4_{-3.1}^{+3.6}$ \\\\
    \hline
    \end{tabular}
        \caption{Best-fitting values for the free parameters adopted in the fit of single cluster pairs (see section~\ref{sec:single_pairs}). For each cluster pair we report the results for the 2gNFW$_{\mathrm{sph}}$+cyl-$\beta$ model. The values of the parameters correspond to the medians of the posterior distributions, with $1\sigma$ errors referring to 16th and 84th percentiles. Subscripts ``1'' and ``2'' refer to the two clusters, respectively, as reported in table~\ref{tab:single_pairs} and figure~\ref{fig:single_pairs}, while subscript ``fil'' refers to the bridge component.}
    \label{tab:spfparams}

\end{table}


\newpage

\section{Corner plots for the stacking fit}
\begin{figure}
    \centering
    \includegraphics[width=\textwidth]{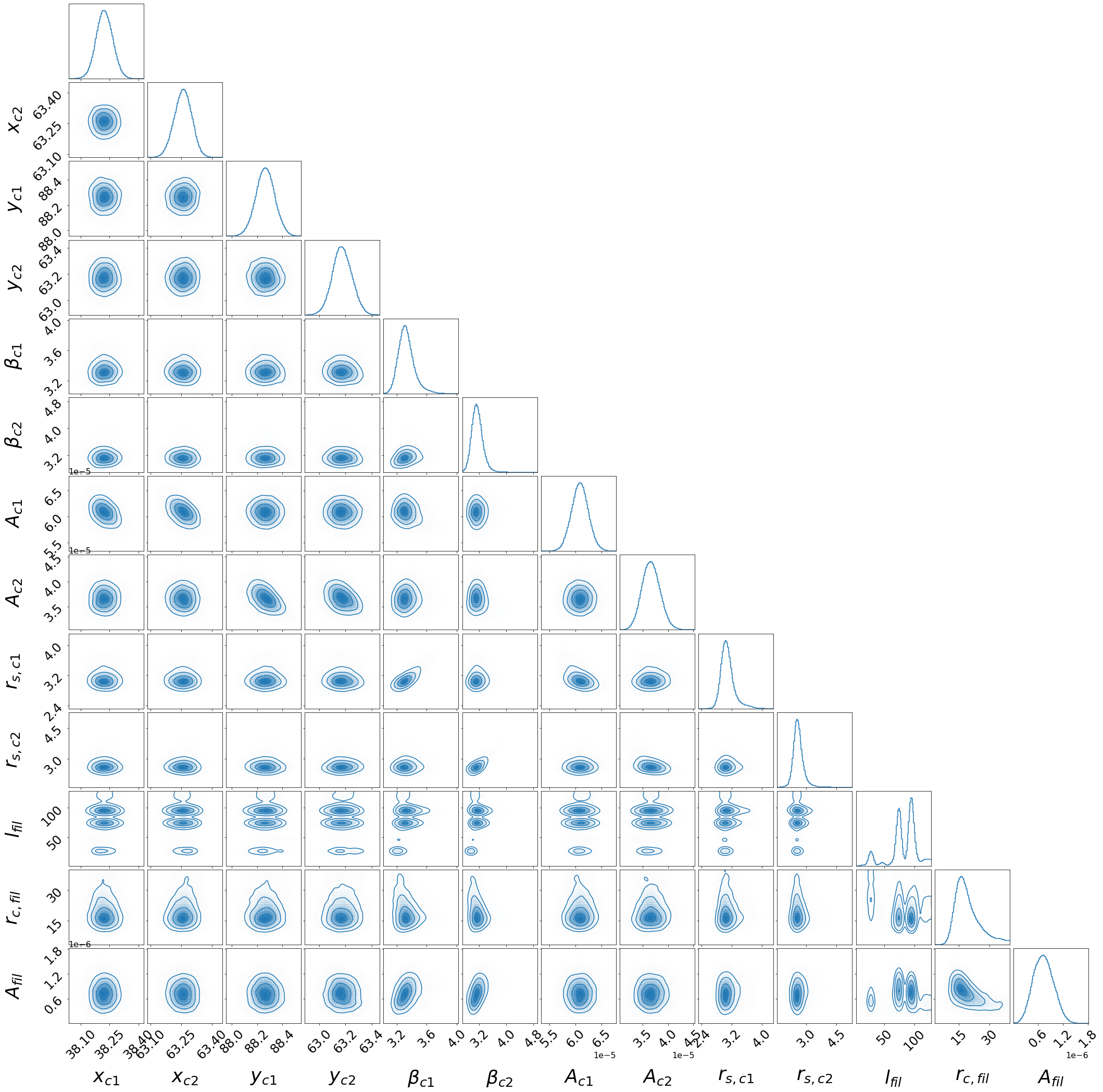}
    \caption{Corner plots for the stack. All units dimensions of length are in pixels. Note the multi-moded posterior for the $l_{\mathrm{fil}}$ parameter.}
    \label{fig:cornerplot_stack}
\end{figure}

In this appendix we show the corner plots for the stacking fit. In figure~\ref{fig:cornerplot_stack}, we overplot the corner plots for the 2gNFW and 2gNFW$_{\mathrm{sph}}$+cyl-$\beta$ models. Note the multi-moded posterior for the parameter $l_{\mathrm{fil}}$. 

\section{Corner plots of simulations fits}
\label{apx:sims_fit_results}

\begin{figure}
    \centering
    \includegraphics[width=\textwidth]{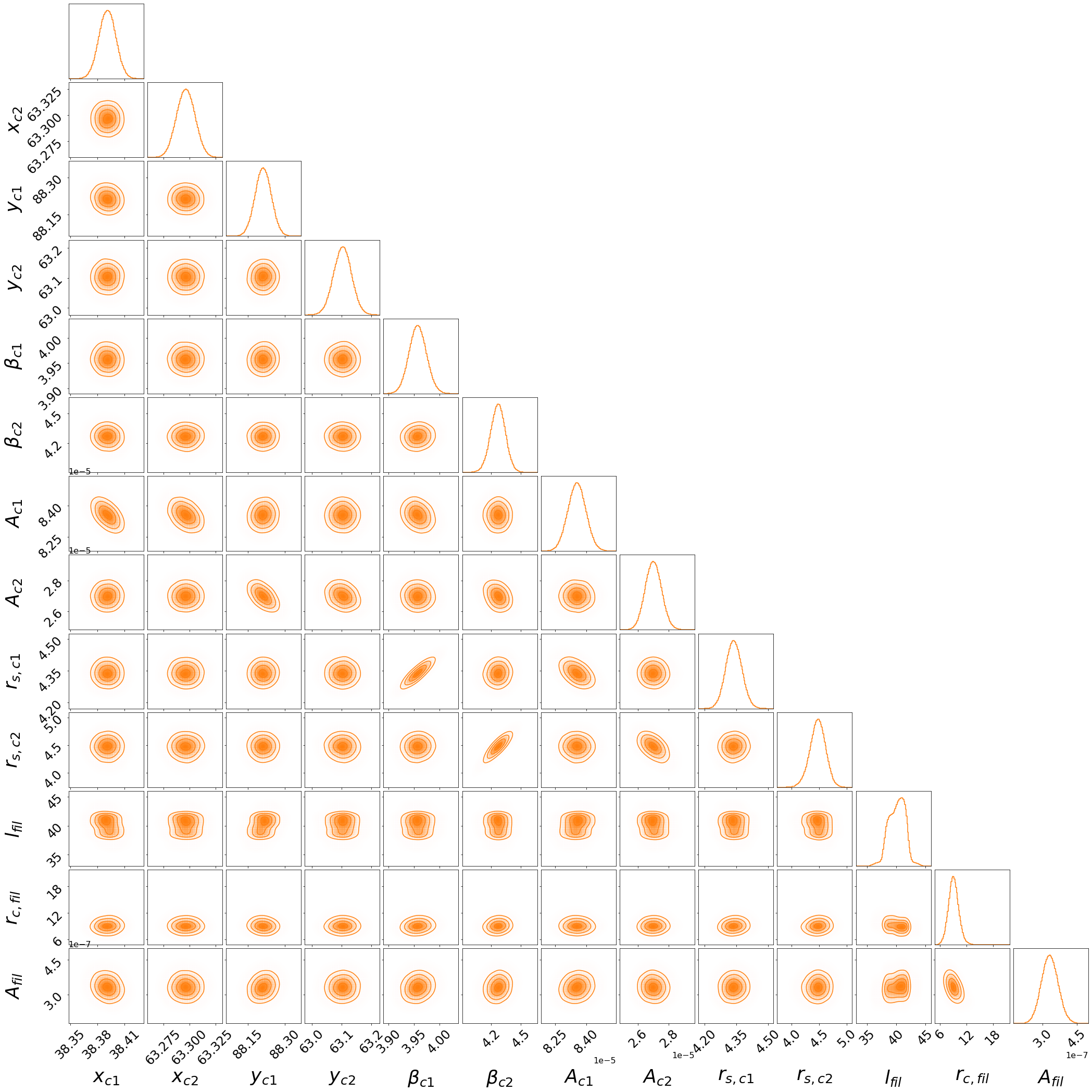}
    \caption{Corner plots for \thethree simulation stack. All units with dimensions of length are in pixels. }
    \label{fig:THE300CornerPlot}
\end{figure}
\begin{figure}
    \centering
    \includegraphics[width=\textwidth]{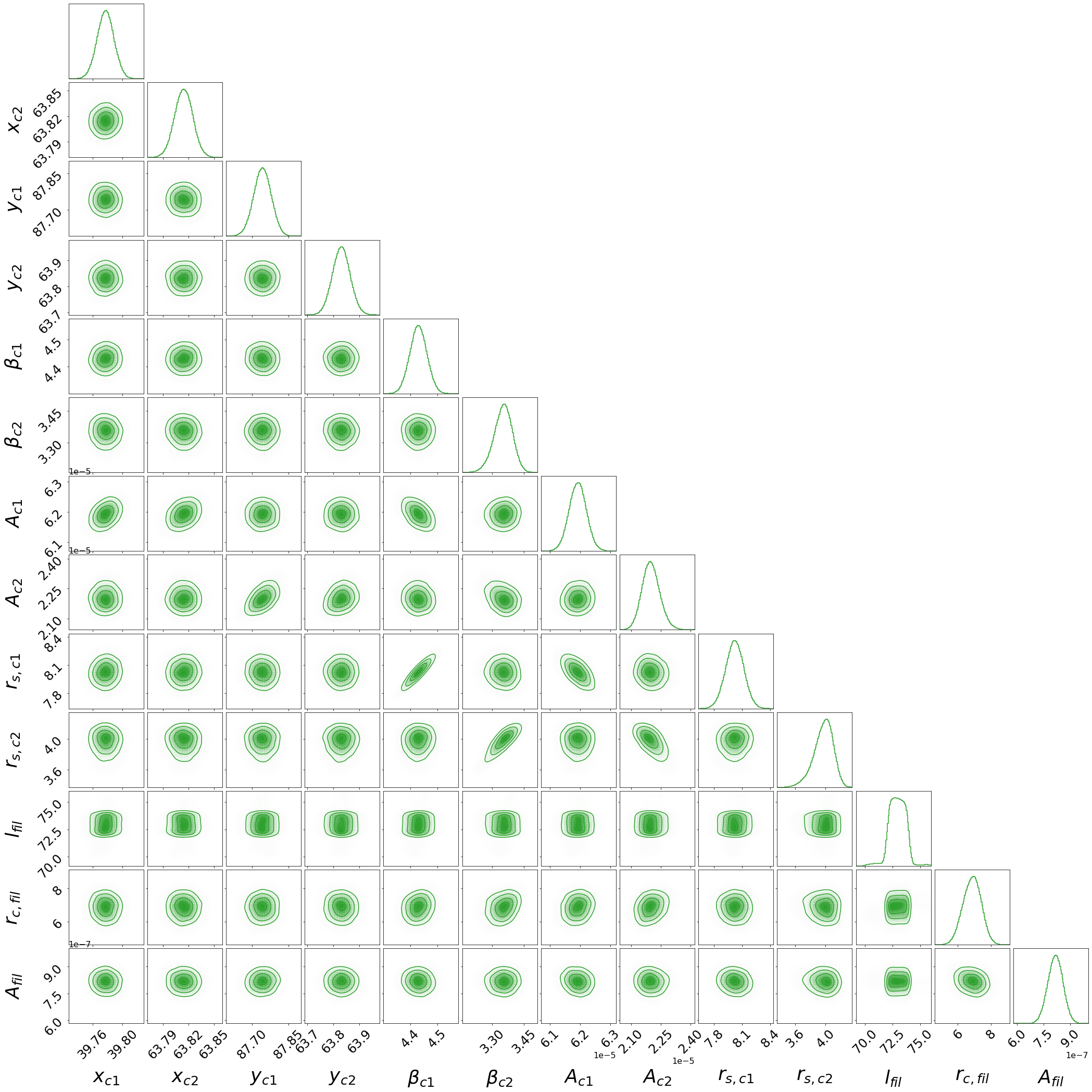}
    \caption{Corner plots for the Magneticum-Pathfinder simulation stack. All units are in pixels. }
    \label{fig:MagneticumCornerPlot}
\end{figure}

In this appendix we report the corner plots of the MCMC fit on simulations in figures~\ref{fig:THE300CornerPlot} and  \ref{fig:MagneticumCornerPlot}.

\end{document}